\documentclass{aa} 

\usepackage{graphicx}

\newcommand{\ten}[1] {10$^{#1}$}

\newcommand{\kms}{km\,s$^{-1}$}
\newcommand{\etal}{et~al.}
\newcommand{\Ha}{\hbox{H$\alpha$}}
\newcommand{\Haw}{\hbox{H$\alpha$\,$\lambda $6563}}

\newcommand{\NIIwa}{\hbox{[N\,{\sc ii}]$\lambda $6548}}

\newcommand{\Msun}{$\mbox{M}_{\sun}$}      
\newcommand{\M}{M~31\ }
\newcommand{\MP}{M~31}
\newcommand{\oasis}{{\tt OASIS}}
\newcommand{\hst}{{\tt HST}}
\newcommand{\stis}{{\tt STIS}}
\newcommand{\sis}{{\tt SIS}}
\newcommand{\foc}{{\tt FOC}}
\newcommand{\wfpc}{{\tt WFPC2}}

\newcommand{\cfht}{{\tt CFHT}}

\newcommand{\aap}{A\&A}                    
\newcommand{\aaps}{A\&AS}                  

\newcommand{\aj}{AJ}                    

\newcommand{\apj}{ApJ}                    
\newcommand{\apjl}{ApJ Lett.}             

\newcommand{\mn}{MNRAS}                   

\newcommand{\pasp}{PASP}                          

\begin{document}

\thesaurus{%
  11              
  (11.09.1 M 31;  
  11.11.1;        
  11.14.1;         
  11.16.1;         
  02.09.1         
  )}

\title{The \M double nucleus probed with \oasis\thanks{
Based on observations collected at the Canada-France-Hawaii Telescope,
operated by the National Research Council of Canada, the Centre National de la
Recherche Scientifique of France, and the University of Hawaii}
and {\tt HST}}
\subtitle{A natural $m=1$ mode?}

\author{R. Bacon\inst{1},
        E. Emsellem\inst{1},
        F. Combes\inst{2},
        Y. Copin\inst{1,5},
        G. Monnet\inst{3} and
        P. Martin\inst{4}
}

\authorrunning{R. Bacon \etal}

\institute{Centre de Recherche Astronomique de Lyon,
           Observatoire de Lyon, 69561 Saint-Genis-Laval Cedex, France
           \and
           DEMIRM, Observatoire de Paris, 61, Avenue de l'Observatoire, 75014 Paris, France
           \and
           ESO, Karl-Schwarzschild-Strasse 2, 85748 Garching, Germany
           \and
           Canada-France-Hawaii Telescope, P.O. Box 1597, Kamuela, HI 96743, USA
	   \and
           Sterrewacht Leiden, Niels Bohrweg 2, 2333 CA, Leiden, The Netherlands 
}

\date{Submitted 27th of Oct.}

\maketitle

\begin{abstract}
	We present observations with the adaptive optics assisted integral field spectrograph 
	\oasis\ of the \M double nucleus in the
	spectral domain around the Calcium triplet at a spatial resolution better than $0\farcs5$~FWHM.
	These data are used to derive the two-dimensional stellar kinematics
	within the central 2\arcsec. Archival \wfpc/\hst\ images in the F300W, F555W and F814W bands
	are revisited to perform a photometric decomposition of the nuclear region.
	We also present  \stis/\hst\ kinematics obtained from the archive.
	
	The luminosity distribution of the central region is well separated into
	the respective contributions of the bulge, the nucleus including P1 and P2,
	and the so-called UV peak.
	We then show, using the \oasis\ kinematical maps, that the axis joining P1 and P2,
	the two local surface brightness maxima, does not coincide with the
	kinematic major-axis, which is also the major-axis of the nuclear
	isophotes (excluding P1). We also confirm that the velocity dispersion
  	peak is offset by $\sim 0\farcs2$ from the UV peak, assumed to mark
	the location of the supermassive black hole.
	The newly reduced {\tt \stis/\hst} velocity and dispersion profiles are then compared 
	to \oasis\ and other published kinematics. We find significant offsets with
	previously published data. Simple parametric models
	are then built to successfully reconcile all the available kinematics.

	We finally interpret the observations using new N-body simulations.
	The nearly keplerian nuclear disk of M31 is subject to
	a natural $m=1$ mode, with a very slow pattern speed (3~km/s/pc for
			$M_{\rm BH} = 7\, 10^7$~\Msun), that 
	can be maintained during more than a thousand dynamical times.
	The resulting morphology and kinematics of the mode can
	reproduce the M~31 nuclear-disk photometry and mean stellar velocity, 
	including the observed asymmetries. It requires a central mass
	concentration and a cold disk system representing between
	20 and 40\% of its mass. Such a slow mode could be excited when 
	interstellar clouds from the more external gaseous disk infall towards 
	the centre.  Nuclear disks formed from accreted gas are
	possible candidates for the precursors of these types of structure, and may
	be common in central regions of galaxies.

\end{abstract}

\section{Introduction}

The role of supermassive black holes (hereafter SBHs) as power sources for active galactic nuclei 
is now widely accepted. The number of SBH candidates is growing rapidly and with more than
20 objects with good black hole mass estimates, 
one can start to search for statistical relationships between 
the black hole mass and the surrounding galaxy (de Zeeuw \cite{Zee00} and references therein). 
Preliminary studies indicate that the SBH mass is correlated with the luminosity
of the host galaxy (Magorrian \etal\ \cite{Mag98}) and, with much less scatter, 
with its velocity dispersion (Ferrarese \& Merritt \cite{Fer00}, Gebhardt \etal\ \cite{Geb00}).
This suggests that the SBH mass is of the order of $10^{-3}$ times 
the mass of the bulges (Merritt \& Ferrarese \cite{Mer00}). 
If this relationship is confirmed, it would have important implications 
on our understanding of the formation of SBHs in galaxies. 
However the statistics is still biased to high SBH mass 
because of observational as well as modeling limitations. 
Extending the study to a larger number of objects and to lower SBH mass 
will require a large observational effort.

Detailed study of individual objects is fundamental not only to better 
constrain the SBH mass, but also to provide observational support on the origin of a possible
relation between the SBH and global properties of the host galaxy. Theoretical work suggests that 
SBHs may influence the central morphology of galaxies through scattering of centrophilic stars
(see e.g. the review by Merritt \cite{Mer99}). 
Nuclear bars or spirals have been observed in the central
region of galaxies, and thus proposed as a triggering mechanism for
the growth of SBHs. Although this does not solve the problem of driving the dissipative
component within the central parsec, it strongly suggests that the hypothesis of
axisymmetry and steady-state widely used to model galaxy cores and constrain the SBH mass may 
not be adequate.

In this paper we will focus our attention on the nucleus of \MP, which is an ideal case for such a 
detailed study: it is well resolved with a diameter of approximately 4\arcsec\ 
($\sim 15$~pc), even at ground-based spatial resolution, it is bright ($m_V \sim 13$), 
does not suffer from dust absorption (contrarily to the Galactic centre) and 
has a favourable inclination ($i \sim 70\degr$).
There is a long history of kinematical observations of this nucleus, with first evidence
for a $\sim10^7$~\Msun\ central SBH by Kormendy (\cite{Kor88}) and Dressler \& Richstone 
(\cite{DR88}). It stood out then as one of the most convincing cases, 
since it was inferred from stellar kinematics obtained from
high signal-to-noise ratio spectroscopic observation at 4~pc ($\sim 1\arcsec$) resolution.
However the modeling was based on the unrealistic assumption of spherical symmetry. Upgrading the
modeling to an axisymmetric (or triaxial) geometry would have been possible, but it was already
known since the Stratoscope~II observations (Light \etal\ \cite{LDS74}), that the light distribution
was not even symmetric. This asymmetry was resolved in an intriguing double 
structure by Lauer~\etal~(\cite{Lau93}) using the \hst\ (pre-COSTAR) {\tt WFPC} imaging capabilities.
In the \hst\ original and post-COSTAR images (Lauer \etal\ \cite{Lau98}), 
the double nucleus appears as a bright 
peak (P1) offset by $\sim0\farcs5$ from a secondary fainter peak (P2), nearly coinciding with the
bulge photometric centre, and the suggested location of the SBH inferred from the previous
spectroscopic observations (see the discussion in Kormendy \& Bender \cite{KB99}). 
\hst\ images from the far-UV to near-IR (King \etal\ \cite{Kin95}; Davidge \etal\ \cite{Dav97}) 
and long-slit spectra (Kormendy \& Bender \cite{KB99}) all
demonstrate that P1 has the same stellar population as the rest of the 
nucleus, and that a nearly point-like 
source produces a UV excess close to P2 (King \etal\ \cite{Kin95}).

The first available two-dimensional maps of the 
stellar kinematics of \MP, obtained by Bacon \etal\ (\cite{Bac94}) 
with the TIGER integral field spectrograph, showed another unexpected feature: 
while the stellar velocity field is roughly centred on P2, 
the peak in the velocity dispersion map is offset by 
$\sim0\farcs7$ on the anti-P1 side. Further
\hst\ spectrographic observations were conducted with {\tt FOS}, as well as
\foc\ (Statler \etal\ \cite{Sta99b}).
In the following we will solely refer to the long-slit \foc\ observations of
Statler \etal\ (\cite{Sta99b}) as they are the only published data. At the \hst\ resolution,
the velocity curve presents a strong gradient and the zero velocity point is offset 
by $\sim 0\farcs16$ from P2 towards P1. 
The velocity dispersion peak reaches a value of $\sigma$ with $440\pm70$~\kms\
(to be compared with the best corresponding ground-based value of $\sim 248$~\kms, 
Kormendy \& Bender \cite{KB99}). 
In the Statler \etal\ (\cite{Sta99b}) data, 
the dispersion peak is nearly centred on P2 (within 0\farcs06).
The \foc\ measurements however suffer from a low signal-to-noise ratio (S/N hereafter)
($\sim 14$ per pixel at P1/P2), and were obtained 
via a complex data reduction procedure to palliate some calibration problems. 
The shapes of the velocity and velocity dispersion curves should thus 
be confirmed with better S/N data.  Such data have been obtained by 
Kormendy \& Bender (\cite{KB99}) using the \sis/\cfht\ spectrograph 
although with a significantly lower spatial resolution of $0\farcs64$ FWHM. 
They confirmed that the dispersion peak
is offset by $\sim 0\farcs2$ (in the direction opposite to P1) from their assumed velocity centre,
and that the nucleus is significantly colder than the surrounding bulge. 

On the modeling and theoretical fronts, the situation remains uncertain. 
There is a consensus regarding the existence of a SBH of a few $10^7$~\Msun\ although
published values range from 3 to $10\,10^7$~\Msun.
Such a black hole mass is needed to explain the 
high value of the velocity dispersion and the fast rotating stellar disk. 
Note however that the precise SBH mass cannot be accurately derived 
since we do not yet have a self-consistent model that can account for the observed properties.
The SBH is assumed to coincide with the centre of the UV peak,
near P2, and possibly with the recently uncovered hard X-ray emission detected by Chandra
(Garcia \etal\ \cite{Gar00}). In that scheme the position of 
the velocity dispersion peak should coincide with the location of the
SBH, roughly consistent with the kinematics presented by Statler \etal\ (\cite{Sta99b}), 
but not with the ground-based
observations of Bacon \etal\ (\cite{Bac94}) and Kormendy \& Bender (\cite{KB99}).

The nature of P1 and the observed dynamical pecularities remain a puzzle. While various
possibilities have been discussed, only two hypotheses have been studied in some detail.
Tremaine (\cite{Tre95}) first proposed a model where an eccentric disk of stars orbiting the SBH
at P2 produced the observed accumulation of light at P1. This parametric model was able to reproduce
the light distribution and some of the kinematical features. Emsellem \& Combes (\cite{EC97})
built the first (and still only) self-consistent models for the nucleus of M~31, via N-body simulations.
They performed simulations of a stellar cluster captured in the SBH potential and 
showed that the available observational properties could 
be reproduced well with properly tuned orbital elements for
the falling cluster. However the timescale for the disruption of the cluster is quite small
($\sim$ \ten{5} years), significantly weakening this scenario. More problematic is the observed 
homogeneity in the colours of P2 and P1 whose explanation would require 
some more fine tuning of the stellar population of the cluster.

As emphasized by Statler \etal\ (\cite{Sta99b}) and Kormendy \& Bender (\cite{KB99}),
the Tremaine (\cite{Tre95}) model then remained the only attractive solution:
it naturally predicts the same stellar population for P1 
and the rest of the nucleus and seems to be, at least qualitatively, compatible with
the observed kinematics. Statler~(\cite{Sta99a}) noted that
it may also help to explain a wiggle observed in the \foc\ velocity 
profile near the location of P1. 
In Tremaine's original paper, some preliminary suggestions were given for
the possible formation of such an eccentric disk. These ideas
have yet to be confirmed by more realistic self-consistent models. 
Such an investigation is also important to determine 
if eccentric disks are common in SBH environments.

The goal of this paper is twofold: (i) add more observational constraints to the 
photometry and kinematics of \M nucleus and eliminate the uncertainty regarding
the location of the velocity dispersion peak;
(ii) investigate Tremaine's model or alternative solutions in more detail.

To achieve these goals, we have performed new observations with the adaptive optics 
assisted integral field spectrograph \oasis\ (Section~\ref{sec:oasis}).
Although these new kinematical maps have a factor two better spatial resolution than previous 
two-dimensional spectroscopic data (Bacon \etal\ \cite{Bac94}),
they cannot compete with the \hst\ spatial resolution. 
The relative centring and orientation of the various key features observed in 
the nucleus are also critical. We therefore performed 
a detailed analysis of the photometry using archival 
\hst/\wfpc\ images (Sect.~\ref{sec:photometry}). 
We thus combined them with new velocity and velocity 
dispersion profiles derived from the recently released \stis/\hst\ data (Section~\ref{sec:stis}).
Together, these data provide new constraints on the dynamics of \M 
(Sect.~\ref{sec:results}). 
We finally interpret the observations in terms of slow $m=1$ modes
and present new N-body simulations to support this scenario
in Section~\ref{sec:simu}. Conclusions are given in Sect.~\ref{sec:conclusion}.

\section{\hst/\wfpc\ photometry}
\label{sec:photometry}

The nucleus of \M is a photometrically
and dynamically decoupled stellar system with respect to the surrounding bulge: it
clearly stands out and shows no clear colour gradient from the UV to 1~$\mu$m 
except for a marginally resolved central UV excess interpreted as a compact
stellar cluster (see e.g. Lauer \etal\ \cite{Lau98}). 
The detailed understanding of these observed structures 
first requires us to reconcile the miscellaneous published photometric and
kinematical data sets. In this context, the \wfpc/\hst\ images are keystones
in the study of the morphology of the nucleus of \MP. These data are now publicly available in a number of 
different bands. Photometric decomposition was done
by Kormendy \& Bender (\cite{KB99}, hereafter KB99) and Lauer \etal\ (\cite{Lau98}). 
We however decided to revisit the surface brightness decomposition 
to provide new quantitative estimates regarding the 
respective contribution and colours of the bulge, the nucleus and the UV peak.
An accurate centring of these structures is critical
for the kinematical study conducted in the following Sections.

\subsection{The data}

We used \wfpc\ images retrieved from the ST/ECF archives in the three following bands\footnote{We also
made use of the F568N exposures (PI Bohlin, 5121): see Appendix~A.}:
F300W, F555W and F814W (PI Westphal, 5236). The individual PC2 exposures 
were combined via a drizzling technique ({\tt drizzle2}) available under {\tt IRAF} 
and implemented by H.-M. Adorf and R. Hook. This led to a pixel size
of $0\farcs022775$, half a standard PC2 pixel. This had the advantage of correcting
for the spatial distortions and partially solve the undersampling problem. 
The drizzling routine was applied to all images using a new pixel fraction
(set with the ``pixfrac'' parameter) of 0.6, except for the F300W exposure for which we kept 
a value of 1 as the individual PC2 images were not dithered.
The output Point Spread Function (hereafter PSF) is roughly a quadratic sum of three contributions:
the optical PSF, the input pixel size and the ''pixfrac'' parameter.
The shorter wavelength of the F300W image thus compensates for the higher ''pixfrac'' value
and gives an output PSF width comparable to the F555W drizzled image
(about 1.6 PC2 pixels). All images were carefully centred, normalised 
using the most recent {\tt PHOTFLAM} values to convert the F300W, F555W and F814W 
into the {\tt VEGAMAG} system. In the following we will simply
write of the $U$, $V$ and $I$ images instead of the F300W, F555W and F814W respectively.
We finally deconvolved these images with the corresponding PSFs
(using a Lucy-Richardson algorithm available in {\tt IRAF}), and reconvolved
them to a common resolution of about $0\farcs06$ arcsecond, using a simple
gaussian PSF. We checked that the residual differences in the final PSFs 
were small and do not affect the results presented below.

\subsection{Decomposition: the bulge, the nucleus and the UV peak}
\label{sec:decomp}

As in Bacon \etal\ (\cite{Bac94}), we used the Multi-Gaussian Expansion formalism 
(Monnet \etal\ \cite{Mon92}, Emsellem \etal\ \cite{Ems94}) to 
build a two-dimensional fit of the surface brightness distribution in the
centre of \MP. We first obtained an excellent fit of the bulge light 
in the $I$ band using 3 Gaussians.
This model is only intended to fit the bulge surface brightness in the PC field.
We also assumed a flat central brightness profile for the bulge by masking
out the central 2\arcsec\ (where the nuclear light dominates) during the fitting process.
The resulting parameters are presented in Table~\ref{tab:MGEbulge}.
\begin{table}
\caption[]{Parameters for the MGE model of the bulge surface brightness in the PC field ($I$ band).}
\begin{center}
\begin{tabular}{l|rrrr}
\hline
\# & $I_i$ & $\sigma_i$ & $q$ & PA \\
 & [L$_{\odot}$.pc$^{-2}$] & [arcsec] & & [deg] \\
\hline
1 & 11012.11  &   3.67  &   1.000 & 45.59 \\
2 &  7709.34  &   8.61  &   0.667 & 45.59 \\
3 & 22970.38  &  26.16  &   0.964 & 45.59 \\
\hline
\end{tabular}
\end{center}
\label{tab:MGEbulge}
\end{table}
\begin{figure}
\centering
\resizebox{8.2cm}{!}{\includegraphics[clip]{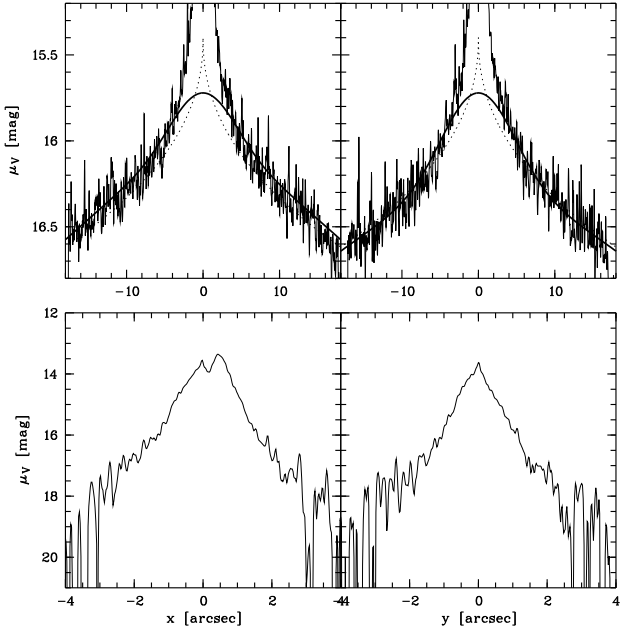}}
\caption{Upper panels: major- (left) and minor-axis (right) $V$ band surface brightness profiles 
with two models: our MGE fit (solid line), and KB99's spherical model (dotted line).
The lower panels show the corresponding residuals after subtraction of the
MGE bulge model.}
\label{fig:bulgecomp}
\end{figure}

\begin{figure}
\centering
\resizebox{8.2cm}{!}{\includegraphics[clip]{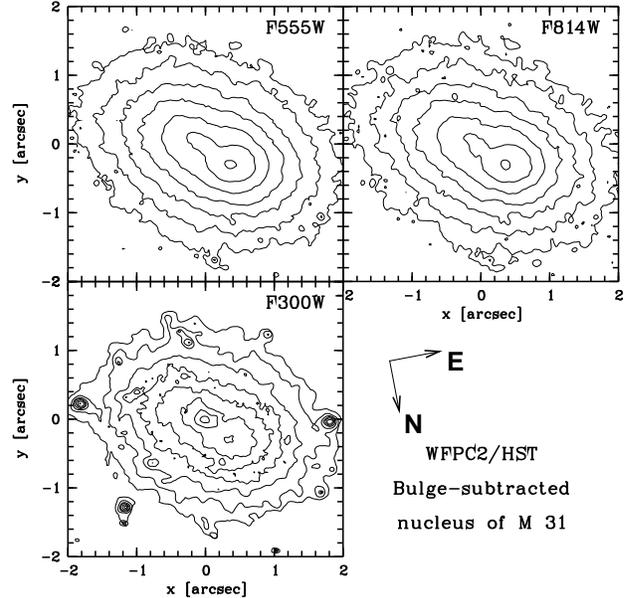}}
\caption{Bulge-subtracted isophotes of the nucleus of M~31 from the \wfpc/\hst\ images
(top right: F814W $I$; top left: F555W $V$ and bottom right: F300W $U$). 
The isophote step is 0.5 magnitude, and the faintest one is 15, 16.3 
and 18.66 for $I$, $V$ and $U$ respectively. The F300W faintest isophotes
are significantly disturbed due to the lower S/N and the presence of blue stars.}
\label{fig:nuc}
\end{figure}

\begin{figure}
\centering
\resizebox{8.2cm}{!}{\includegraphics[clip]{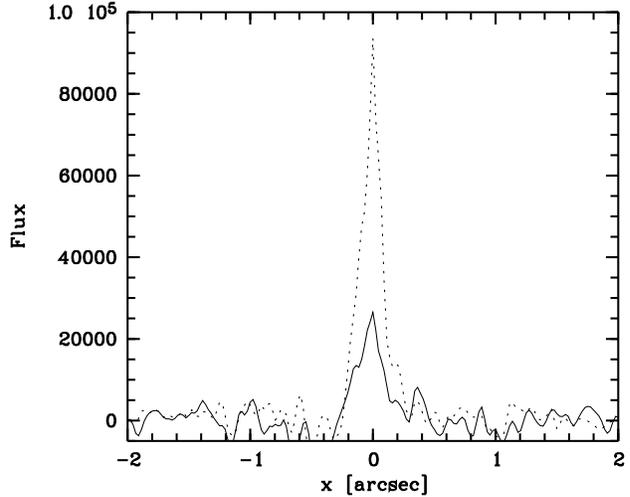}}
\caption{Deconvolved bulge$+$nucleus subtracted surface brightness profiles of the so-called ``UV peak'' in the
 nucleus of M~31 (\wfpc/\hst) in $V$ (solid line) and $U$ (dashed line).
Flux units are in $L_{\odot}$.pc$^{-2}$ in the corresponding band
(corrected for galactic extinction).}
\label{fig:peakUV}
\end{figure}

Our bulge representation differs from KB99's as the former has 
a non-zero and varying ellipticity, and a surface brightness profile 
which flattens towards the centre.
The contribution of the bulge inside 2\arcsec\ is in fact impossible to quantify
given the present data and should therefore be taken as an unknown free parameter of the model.
We then normalised the $I$ band bulge model to the $U$ and $V$ bands 
(see the comparison with KB99's model in Fig.~\ref{fig:bulgecomp}) and subtracted
its contribution, providing bulge-subtracted images (see Fig.~\ref{fig:nuc}). 
Assuming that the UV peak detected by King \etal\ (\cite{Kin95}) has 
a negligible contribution in the 
$I$ band, we normalised the nuclear $I$ band image
and subtracted it from the (bulge-subtracted) $U$ and $V$ images (Fig.~\ref{fig:peakUV}).

It is surprising to see how well this simple decomposition
procedure works: the only significant residuals in the $U$ band indeed come from the UV peak 
(also seen in the $V$ band) and from very blue (extreme horizontal branch?)
stars clustering around the nucleus (Brown \etal\ \cite{Brown98}). 
This shows that our bulge model is adequate (keeping in mind
that the exact bulge light contribution remains unknown in the central arcsecond).
It accounts for the observed ellipticity and position angle of the
isophotes in the central 20 arcseconds. It also confirms that the core of
\M can indeed be decomposed into three distinct components: the bulge, the nucleus,
including P1, and the UV peak.
The obtained colours for the bulge and the nucleus are then (corrected for
galactic extinction: $0.42$, $0.24$, and $0.1436$ mag in $U$, $V$ and $I$ respectively): 
$(V-I)_B = 1.26$, $(U-V)_B = 1.98$ and $(V-I)_N = 1.30$, $(U-V)_N = 2.36$.
The UV peak has an average axis ratio of $\sim 0.7 \pm 0.05$, a PA of $\sim 62\degr \pm 8$,
a major-axis FWHM of $0\farcs21$, and has $(U-V) = -0.35$.
We therefore find that the nucleus is redder than the bulge:
this is in contradiction with the results of Lauer \etal\ (\cite{Lau98}) who found
that the nuclear region is bluer than the bulge in the $U - V$ colour.
We checked and confirm the robustness of this result by reexamining the original images.
We thus cannot explain this discrepancy even considering the slight differences in the
reduction process. 

\subsection{Centres and positioning}
\label{sec:centre}

Following the photometric decomposition achieved in Sect.~\ref{sec:decomp},
we have chosen our reference centre, position $[0,0]$ throughout the paper,
to correspond to the central UV maximum in the F300W image (see Fig.~\ref{fig:peakUV}).
This choice was motivated by the fact that the UV peak is very bright
and just resolved in the \wfpc\ images, allowing a very accurate
determination of its centre. It is also thought to correspond
to the true position of the presumed central massive black hole
as discussed by KB99. KB99 defined their spatial zero radius $x_{\mbox{\tiny KB}}$ to be the velocity centre
at the resolution of their SIS data. Comparing their surface brightness
profile (their Fig.~8) with the F300W and F555W \wfpc\ photometry, we find that $x_{\mbox{\tiny KB}}$
is $0\farcs031$ away (towards P1) from the position of the UV peak (our $[0,0]$).
KB99 quoted a value of $0\farcs051 \pm 0\farcs014$ between the UV peak
and $x_{\mbox{\tiny KB}}$. This difference (of $0\farcs02$) is easily 
accounted by the fact that the contribution of the ``UV peak'' 
quickly decreases at longer wavelengths, the local maximum around P2
shifting away from P1, with an offset of
about $0\farcs023$ already in the V band (roughly one dithered pixel).
Throughout this paper, P2 is defined as the secondary surface brightness maximum
in the $I$ band, where we assume that the UV excess has a negligible contribution.
P2 is indeed not coincident with the location of the UV peak: it
shows up (in the $I$ band) as an elongated structure, it centre being at 
about $0\farcs076$ from the UV peak (see Fig.~\ref{fig:orient}).

The spatial zero radius of Statler \etal\ (\cite{Sta99b}, Sta+99) is 
about $-0\farcs025$ from the UV peak (away from P1), 
according to their approximate surface brightness long-slit profile 
(spectral domain from $\sim 4000$ to 5450~\AA, see their Fig.~3). Their zero radius
is therefore $0\farcs025 + 0\farcs031 = 0\farcs056$ from the one defined in KB99.
However, careful inspection of Fig.~6 in Sta+99
shows that the kinematical data of KB99 has been uncorrectly shifted by $\sim 0\farcs13$
(towards P1) in order to be consistent with their FOC kinematics: this shift is then 
$\sim 0\farcs074$ larger than it should be. The fact that Sta+99
managed to roughly fit the kinematical data of KB99 comes partly from 
the way the extrapolation of their one-dimensional $V$ and $\sigma$ FOC 
profiles was achieved, assuming almost no dependency perpendicular to 
the slit (see Sect.~5.1 of Sta+99), and partly from this uncorrect spatial registering.

To summarize, the zero radius defined in Sta+99 
and in KB99 are offset by $-0\farcs025$ and $0\farcs031$, respectively, with respect to our
spatial zero point taken as the location of the UV peak in the F300W band
(Fig.~\ref{fig:orient}).
We do not include here unknown (small) potential offsets perpendicular to their slits.

\section{\oasis\ spectrographic data}
\label{sec:oasis}

\subsection{Observations}
\label{sec:observations}
We observed the nucleus of \M in December 98 using the 
{\oasis\footnote{\oasis\ stands for {\bf O}ptically {\bf A}daptive {\bf S}ystem for 
{\bf I}maging {\bf S}pectroscopy. It has been funded by the CNRS, the MENRT and 
the R\'egion Rh\^one-Alpes} 
integral field spectrograph attached to the CFHT adaptive focus bonnette ({\tt PUEO}).
\oasis\ is the successor of the {\tt TIGER} integral field spectrograph successfully used at CFHT 
between 1987 and 1996.
The instrument design follows the {\tt TIGER} concept described in Bacon \etal\ (\cite{Bac95}), and
includes a spectrographic mode, in which the spatial sampling is achieved via a microlens array,
and an imaging mode. Various sampling sizes and spectral resolutions are
available, from 0.04 arcsec to 0.16 arcsec per lens, and $400 < R < 4000$ respectively. \oasis\ has 
been designed and built at the Observatoire de Lyon and is operated at CFHT as a guest instrument.

We used the MR3 configuration covering the Ca triplet (8500 \AA) region. This 
configuration was favoured against the classical Mg2 (5200 \AA) wavelength range because {\tt PUEO}, 
like all other adaptive optics systems, 
performs better at longer wavelength (Rigaut \etal\ \cite{Rig98}). 
The selected spatial sampling of $0\farcs11$ per (hexagonal) lens provides a field of view 
of 4$\times$3~arcsec$^2$. Details of the instrumental setup are given in Table~\ref{tab:config}.

A total of nine exposures, each 30 mn long, were obtained during a seven day period
(December 17--24).
The atmospheric conditions were photometric. Seeing
conditions were generally good, but changed rapidly (see Sect.~\ref{sec:psf}).
All exposures were centred on the nucleus, with only small offsets (typically of the order of 1--2 
sampling size). Neon arc lamp exposures were obtained before and
after each object integration, and other required configurations exposures (bias, dome flatfield, 
micropupil) taken during daytime. Bright sky flatfields were also observed at dusk or dawn. 
The star HD 26162 (K2 III) was chosen as a kinematical template, and observed with the same 
instrumental setup.
\begin{table}
  \caption[]{\oasis\ instrumental setup}
  \begin{flushleft}
  \begin{tabular}{ll}
  \hline
  \multicolumn{2}{c}{{\tt PUEO}} \\
  \hline
  Loop mode & automatic \\
  Loop gain & 80 \\
  Beam splitter & I \\
  \hline
  \multicolumn{2}{c}{\oasis} \\
  \hline
  Spatial sampling & 0.11 arcsec \\
  Field of view &  $4 \times 3$ arcsec$^2$ \\
  Number of spectra & 1123 \\
  Spectral sampling & 2.17 \AA\ pixel$^{-1}$ \\
  Instrumental broadening ($\sigma$) & 69~\kms \\
  Wavelength interval & 8351--9147 \AA \\
  \hline
  \end{tabular}
  \end{flushleft}
  \label{tab:config}
\end{table}

\subsection{Data reduction}
\label{sec:datared}

The \oasis\ data were processed using the dedicated XOasis software (version 4.2) developed in 
Lyon\footnote{Software and documentation are available at 
http://www-obs.univ-lyon1.fr/$\sim$oasis/home/home\_oasis\_gb.html}. The standard 
\oasis\ reduction procedure includes CCD preprocessing (bias and dark subtraction), 
spectra extraction, wavelength calibration,
spectro--spatial flatfielding, cosmic rays removal and flux calibration. Given the 
high surface brightness of the nucleus of \MP\ and the small field of view, 
no sky subtraction was required.

\subsubsection{Spectra extraction and calibration}
The extraction of the spectra is the most critical phase of the reduction process. 
It uses a physical model 
of the instrument with finely tuned optical parameters, such as the rotation angle of 
the grating with respect to the CCD column, the tilt of the grism, the collimator and camera 
focal ratios, etc. The free parameters are fitted using a set of calibration exposures
(micropupils -- obtained without the grism --, arc and tungsten continuum frames) and saved 
in an extraction mask, later used to retrieve the precise (subpixel) location of each spectrum. 
Small shifts of the spectra location obtained at different airmass may occur due to
mechanical flexures. These are corrected by including a global shift between the object 
exposure and the extraction mask. This offset is measured via 
the cross-correlation of the arc frame associated with the science exposure,
and the one associated with the reference continuum exposure (used to build the extraction mask). 
Shifts are typically 0.1--0.2 pixel. Given that observations of \M were split 
over eight nights, we built two extraction masks from two sets of calibration exposures obtained
at the start and end of the run respectively. These two masks 
were found to be almost identical, both giving an excellent fit (0.08 pixel RMS residuals
between the model and the continuum exposure). Various extraction tests using
each mask independently showed no significant differences, and we chose to use only the first
mask. All spectra were extracted via an optimal extraction algorithm (Bacon \etal\ \cite{Bac01}) 
applied to each exposure (object and associated arcs), 
finally providing 9 science datacubes with 1123 spectra each. 

These datacubes were then wavelength calibrated using the associated (nearest) arc exposure. 
The goodness of the optical model guarantees that the raw extracted spectra
do already have a good wavelength precalibration.
First order correction polynomials were thus sufficient to finalise the calibration, with
typical RMS residual errors of 0.04--0.05 \AA\ ($\sim 1.5$~\kms). 
After wavelength rebinning, the spectro-spatial
flatfield correction was applied to each individual spectrum. This correction is obtained 
using a combination of a high S/N continuum datacube (tungsten lamp) and a twilight sky flat 
field datacube. The former is used to correct the spectral variations, 
while the latter allows to correct lens-to-lens flux variations. 
Cosmic rays are then detected via a comparison beween each spectrum 
and its neighbours, and removed. The overall fraction of pixels affected by cosmic 
rays is small, typically 0.2--0.3\%. Finally, spectra are truncated to a common wavelength range 
(8361--8929 \AA) covering the Ca absorption line triplet. No flux calibration was performed, 
since this is not required to measure the stellar kinematics.

\subsubsection{PSF determination, and merging of the datacubes}
\label{sec:psf}
PSF determination deserves special attention. Indeed, a precise knowledge of the spatial 
PSF is critical to be able to compare in detail datasets which span a rather large range 
in spatial resolution (from \hst\ to ground-based long-slit observations).

Reconstructed images are computed by direct summation of spectra along 
the wavelength range followed
by an interpolation on a square grid. As expected, the 9 reconstructed images present
significantly different PSFs. We used the available
\hst/\wfpc\ $I$ band images of the \M nucleus (see Sect.~\ref{sec:photometry}) to have 
an accurate estimate
for each of them. Assuming a parametric shape for the PSF,
we performed a simple least-squares fit between the convolved \hst\ image and the 
equivalent reconstructed \oasis\ image. 
Free parameters for the fit also include the relative alignement 
and rotation between the two images, as well as the corresponding flux normalisation factor.
\begin{figure}
\centering
\resizebox{8.2cm}{!}{\includegraphics[clip]{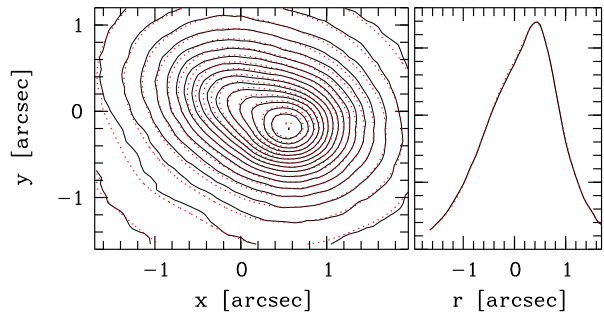}}
\caption{PSF fitting of \oasis\ exposure \#6. Left panel: Contour plot of the reconstructed image (solid line) and the convolved \hst\ image (dashed line). Right panel: photometric major-axis cut.}
\label{fig:psf6}
\end{figure}
\begin{figure}
\centering
\resizebox{8.2cm}{!}{\includegraphics[clip]{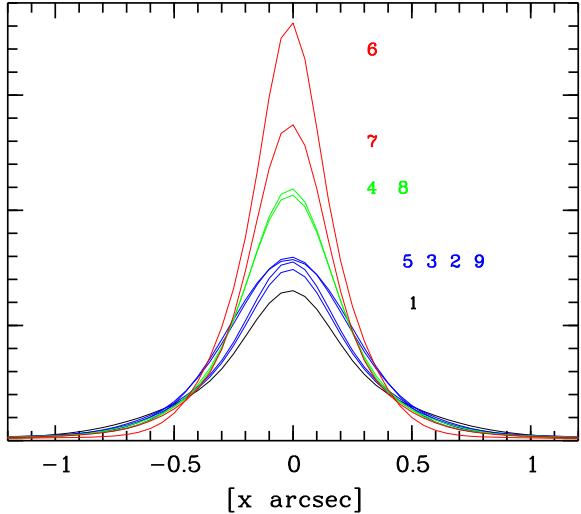}}
\caption{Fitted PSF of the nine \oasis\ exposures. FWHM range from 0\farcs79 (exposure \#1)
to 0\farcs39 (exposure \#6)}
\label{fig:psfcut}
\end{figure}
\begin{table}
  \caption[]{PSF parameters of the merged exposures}
  \begin{flushleft}
  \begin{tabular}{lrrrrrrr}
  \hline
  ID & Exp. & FWHM & $\sigma_1$ & $\sigma_2$ & $I_2/I_1$ & $\sigma_3$ & $I_3/I_1$\\
  \hline
  M2 & 6 -- 7 & 0.41 & 0.15 & 0.29 & 0.98 & 0.448 & 0.023 \\
  M8 & 2 -- 8 & 0.50 & 0.18 & 0.35 & 1.07 & 0.407 & 0.030 \\
  \hline
  \end{tabular}
  \end{flushleft}
  \label{tab:merge}
\end{table}

The reference \hst/\wfpc\ F814W ($I$) band filter covers a wavelength range
which extends well beyond the spectral limits of the \oasis\ datacubes. However, colour gradients
in the \M nucleus are very small in the visible, 
so we decided to neglect the effects due to the differences in the response curve:
we checked that this had a negligible effect on the final PSFs parameters.
The {\tt PUEO} adaptive optics performance heavily depends on the ``original'' 
seeing conditions, the brightness 
of the guiding source and the wavelength range. 
\oasis\ observations of the \M nucleus are thus challenging 
for {\tt PUEO}, given the low contrast and complex structure in the \M nucleus,
and the relatively blue wavelength range of the observations (compared to the more classical
near-infrared H or K bands). Typical {\tt PUEO} PSFs at these 
wavelengths are not diffraction limited, but show a core of a few tenths 
of an arcsec surrounded by a large halo. A sum of $3$ gaussian 
functions provides a reasonable approximation.
An illustration of the quality of the fit is given in Fig.~\ref{fig:psf6}. 
This procedure does not only provide PSF parameters but, 
as mentioned above, also the precise relative centring between the \oasis\ exposures 
and the \hst/\wfpc\ image.

All nine computed PSFs are displayed in Fig.~\ref{fig:psfcut}.
There are large differences in the resolutions of the nine exposures, which
range from 0\farcs79 to 0\farcs39 FWHM. 
Two merged exposures have thus been created using two different sets of exposures 
ranked according to their resolution:
\begin{itemize}
\item  the first one, hereafter called M2, with the highest spatial resolution is the combination of 
exposure \#6 and \#7 only.
\item the second one, hereafter called M8, with a higher S/N but a corresponding lower spatial 
resolution, is the combination of all exposures except exposure \#1.
\end{itemize}
The selected datacubes were merged using the accurate centring given by the PSF fitting procedure.
Interpolation on a common square grid is performed before averaging the data. One lens 
located in the outer part of the nucleus  was affected by a bad CCD column and was discarded before 
interpolation. The PSF fitting was subsequently performed on the merged datacube and the results are 
given in Table~\ref{tab:merge}. The M2 and M8 exposures have cores FWHM of $0\farcs34$
and $0\farcs41$ respectively, both with extended wings giving a global FWHM of $0\farcs41$
and $0\farcs50$ respectively.

\subsubsection{Computation of line-of-sight stellar velocity distributions}
\label{sec:losvd}
A velocity template spectrum was obtained using the reference star HD~26162 datacube, summed 
over an aperture of $1\arcsec$ to maximize the S/N. The spectrum was then continuum divided and 
rebinned in ln($\lambda$). The same procedure was applied to the merged datacubes of \MP. 
We used the Fourier Correlation Quotient (FCQ) method, originally developed by Bender 
(\cite{Bender90}), to derive the kinematical maps.
The kinematics was parametrized using
a simple Gaussian and complemented using third and fourth order Gauss-Hermite
moments (van der Marel \& Franx \cite{vdMF93}).
All velocities were offset to heliocentric values. A systemic velocity of $308$~\kms\ 
was deduced by comparing our data to the data of KB99
whose kinematics extends far enough to observe the 
slow rotation of the  bulge (see Sect.~\ref{sec:results}).

\subsubsection{Bulge subtraction}
\label{sec:bsub}

Since we are mostly interested in the kinematics of the nucleus of \MP, we 
would like to subtract the contribution of the bulge from the \oasis\ merged datacubes.
This was performed using the photometric model 
derived in Sect.~\ref{sec:decomp} (see KB99 for a similar approach). 
We extracted a mean bulge spectrum 
from the outer part of the \oasis\ field, which we fitted making use of 
a library of stellar templates to get a high S/N reference spectrum.
We then used a simple Jeans dynamical model (see Emsellem~\etal~\cite{Ems94})  to derive the 
velocity and velocity dispersion of the bulge (taking into account the instrumental
setup and seeing) in the \oasis\ field of view.
The resulting spectra were finally directly subtracted from the \oasis\ datacubes
(M2 and M8) after the proper luminosity normalisation. 
We checked that slight variations in the continuum
shape or absorption line depths did not produce any significant differences on the final
bulge-subtracted datacube spectra. The results do however weakly depend 
on the details of the dispersion model used for the bulge: 
the same subtraction procedure was therefore also applied
 using a constant dispersion of 150~\kms\  for the bulge. The difference comes
mainly from the higher assigned central dispersion to the bulge in the Jeans model.
In Sect.~\ref{sec:bulgesub}, we will only discuss results 
from the bulge-subtracted M8 \oasis\ datacube, 
which has a significantly better S/N.

\section{\stis/\hst\ data}
\label{sec:stis}
\begin{figure}
\centering
\resizebox{8.2cm}{!}{\includegraphics[clip]{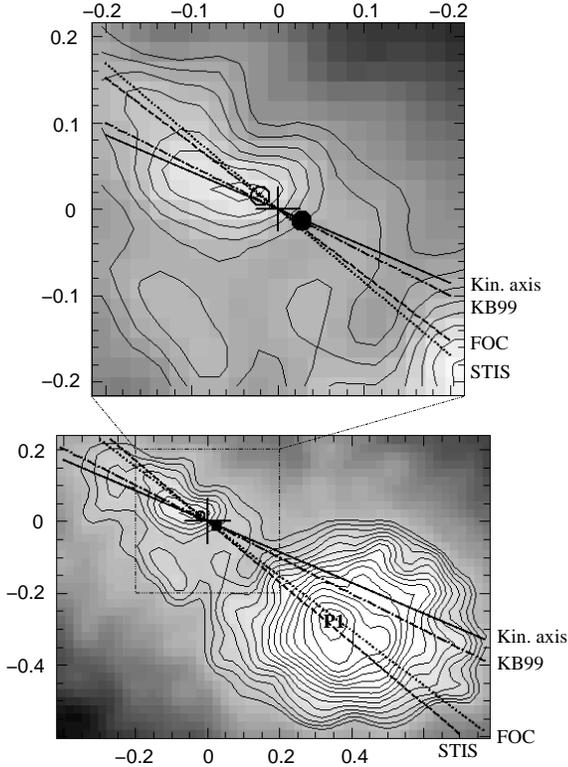}}
\caption{The central region of the deconvolved \wfpc\ image in the $I$ band showing
the location of P1, P2 and the UV peak.
Isophotes are drawn from 11.79 to 12.49 with a step of 0.05~mag$\,$arcsec$^{-2}$.
The UV peak, located at $[0,0]$, is marked by a cross. The solid line
represents the kinematical major-axis, with which P2 elongation is well aligned. 
The slit axis for the {\tt SIS} (KB99), \stis\  (this paper) and \foc\ (Sta+99) 
data are represented by dotted-dashed, dashed and dotted lines, respectively.
The filled and empty circles represent the location of the spatial zero point
of the \sis\ and \foc\ long-slit data, respectively.
The upper panel shows a zoom in the central $0\farcs2$.
Note the distance between the maximum of P1 (marked with the text ''P1'')
and the kinematical major-axis.}
\label{fig:orient}
\end{figure}

The nucleus of M~31 was observed with \stis\ at PA$=39\degr$
in July 1999, using the G750M grating and the $52\times0\farcs1$
slit (Proposal 8018, PI Green). The spatial pixel was $0\farcs05071$,
and the velocity resolution about 38~\kms. 
We have retrieved the 8 corresponding
science exposures and calibration files from the ST/ECF archives 
in this configuration\footnote{We did not include here the G430L exposures
as we were solely interested in the kinematics.}. After the data reduction
provided via the ST/ECF pipeline using the best reference files, we corrected
the individual frames for the residual fringing using the dedicated routines
available under {\tt IRAF} (Goudfrooij \& Christensen \cite{Gou98}) and
the available contemporaneous flat fields. We combined the individual exposures
after careful recentring, and extracted individual spectra, 
which were rebinned in $\ln{(\lambda)}$ for the kinematical measurements. 
The spatial centre of the long-slit luminosity profile was then
accurately determined taking the same
reference zero point as in the \wfpc/\hst\ data (see Sect.~\ref{sec:centre}).
Spectra were binned along the slit to increase the S/N
outside $0\farcs85$ from the centre. We also retrieved \stis\ exposures
from the K0III star HR~7615 (same configuration) from the archive
to be used as a stellar template, and reduced them in the same way.
The line-of-sight velocity distribution at each spatial location
was finally derived using the Fourier Correlation Quotient method
as in Sect.~\ref{sec:losvd}. We estimated a maximum velocity error
of $\sim 8$~\kms\ due to the slit effect using the present \stis\ characteristics 
(see Bacon \etal\ \cite{Bac95}). In the present paper, we will mainly focus 
on the first two moments, namely the velocity and velocity dispersion,
only using the higher order Gauss-Hermite moments to derive an estimate
of the first two true velocity moments. We finally performed the subtraction
of the contribution of the bulge as in Sect.~\ref{sec:bsub}.

\section{Results}
\label{sec:results}

\subsection{\oasis\ stellar velocity maps}
\begin{figure*}
\centering
\resizebox{16cm}{!}{\includegraphics[clip]{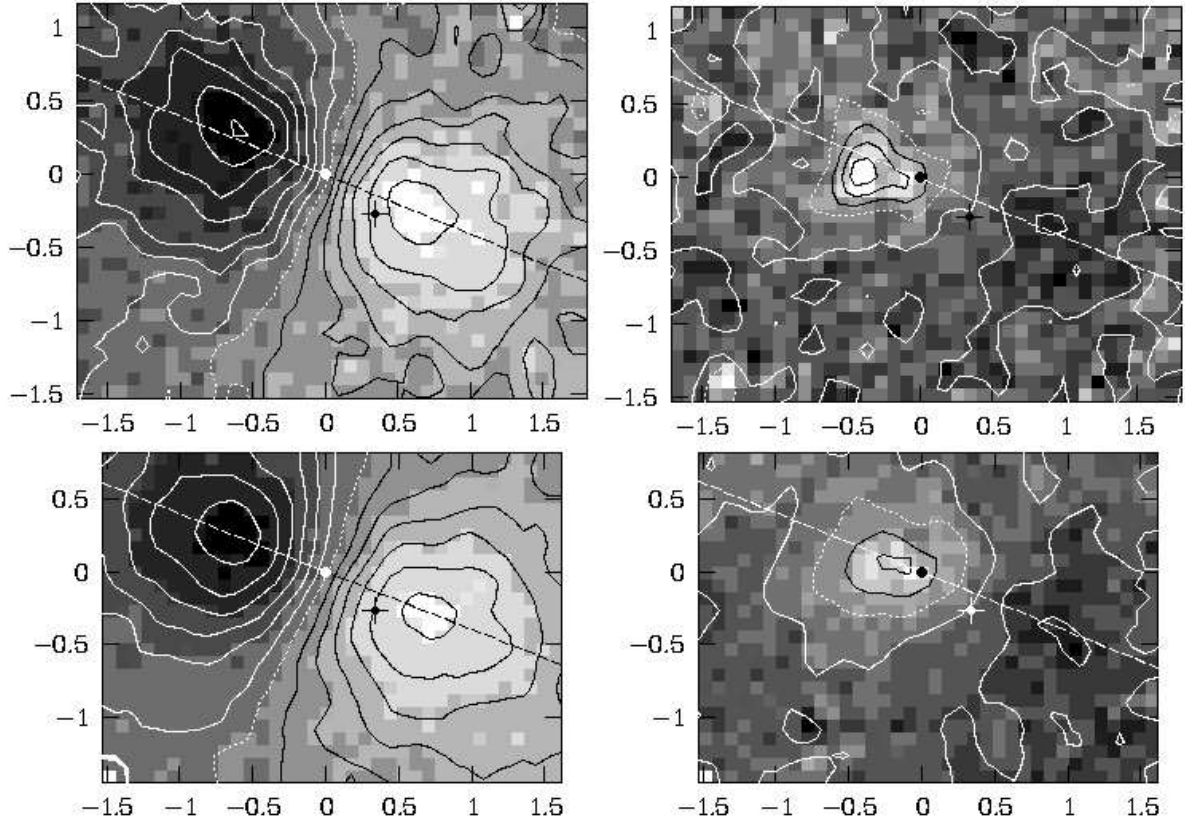}}
\caption{\oasis\ stellar kinematical maps: stellar velocities (left panels)
and dispersions (right panels). The top and bottom panels correspond to 
the high spatial resolution datacube (M2) and the high S/N datacube (M8), 
respectively. 
The dotted white lines display the zero isovelocity contour and the 
200~\kms\  isovelocity dispersion. The step is 25~\kms\  in all panels.
The centre of the UV peak, our $[0,0]$, is marked by a filled circle,
the cross marking the centre of P1.
The dashed lines indicate the kinematical major-axis.
North is at a PA of $192\degr$, East at its right, and the scale is in arcsecond.}
\label{fig:vmap}
\end{figure*}

In Fig.~\ref{fig:vmap}, we present the stellar velocity and velocity 
dispersion maps\footnote{\oasis\ data are available in ascii/fits form at 
	{\tt http://www-obs.univ-lyon1.fr/$\sim$tigerteam/bacon01.html}} 
for the two final datasets M2 and M8. Both give similar results,
as shown in Fig.~\ref{fig:cutm2m8}, although the difference in spatial resolution
can be seen in the central velocity gradient and velocity dispersion peak.

We compare these results with the KB99 kinematical profiles
obtained with the \sis/\cfht\ spectrograph at a spatial resolution of 
$0\farcs65$ FWHM. The \sis\ slit position has been spatially shifted 
by $+0\farcs031$ to account for the small offset between the location of the UV peak 
and the velocity centre as defined by KB99
(see Sect.~\ref{sec:centre}). A simulated \oasis\ slit was computed using the 
surface brightness weighted average of the kinematical quantities ($V$ and $\sqrt{V^2+\sigma^2}$)
over an equivalent slit opening of $0\farcs35$ at PA$= 52.5\degr$ (Fig.~\ref{fig:kor99}). 
The comparison was done using the M8 data set. 
The agreement is excellent, with 8 and 9~\kms\  RMS 
residual in velocity and velocity dispersion, respectively.
Note that the difference due to spatial resolution 
is largely smoothed by the averaging over the 0\farcs35 slit width.

The precise axis and centre of symmetry of the velocity field 
was estimated using a fit with a simple analytical function. 
In that model, the velocity field was parametrized as a sum of first order Gauss-Hermite-like
functions $x' \cdot \sum_i\; \mathrm{Gauss}_i(x',y')$ where $x',y'$ are cartesian coordinates on the sky. 
Free variables are the coordinates of the individual centres ($x_0^i, y_0^i$), 
the principal axis angle ($\theta_0$), 
the Gaussian parameters ($\sigma_i, I_i, q_i$) and a zero velocity. 
Experiments show that the kinematic centre and tilt are insensitive to the details
of the fitting functions provided it is antisymmetric. 
The two merged datacubes gives consistent results. 
A use of only two components provides a reasonable fit with 
16 \kms\  RMS residuals. The centre of symmetry is found at [0,0] 
within 20 mas and the kinematic axis with $\theta_0 = 56.4 \pm 0.2 \degr$. 
The velocity at [0,0] is not zero, but $\sim -9$~\kms.
\begin{figure}
\centering
\resizebox{8.cm}{!}{\includegraphics[clip]{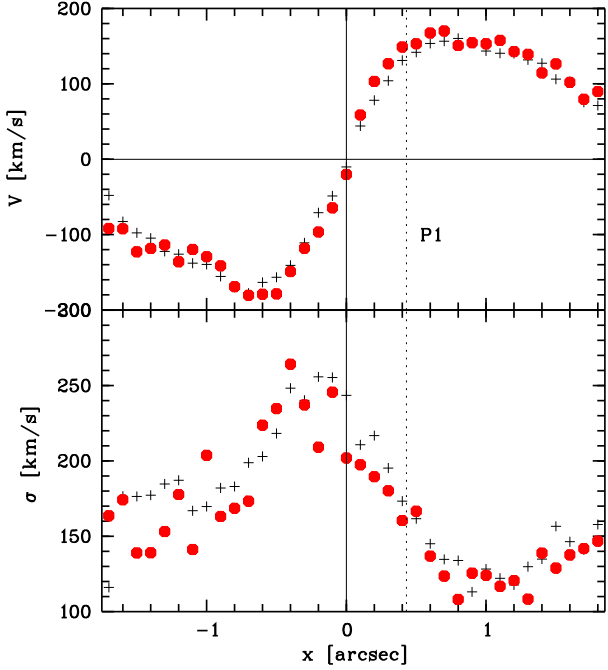}}
\caption{\oasis\ velocity profiles along the kinematic axis (PA$= 56.4\degr$, averaged over a 
width of $0\farcs2$). Top panel: stellar velocity. Bottom panel: stellar velocity 
dispersion. The crosses and circles correspond to the high S/N (M8), and the high spatial resolution 
(M2) datacubes respectively.}
\label{fig:cutm2m8}
\end{figure}
\begin{figure}
\centering
\resizebox{8.cm}{!}{\includegraphics[clip]{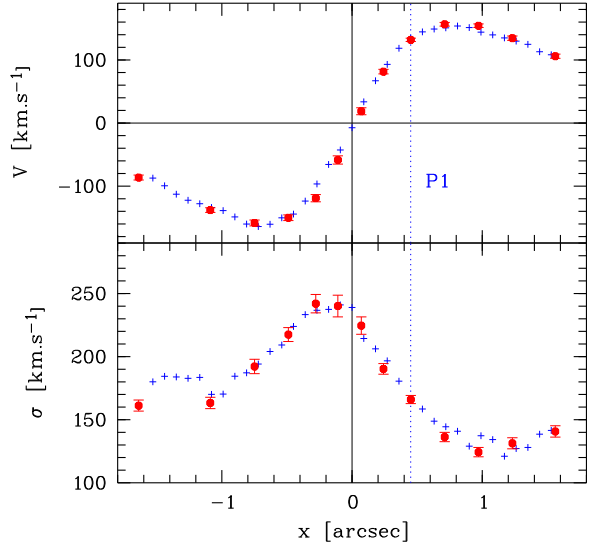}}
\caption{Comparison between the \sis\ kinematics (circles; Kormendy \& Bender
\cite{KB99}; PA$=52.5\degr$, slit width of $0\farcs35$)
and the \oasis\ equivalent slit (crosses). 
Top panel: mean velocity. Bottom panel: velocity dispersion.}
\label{fig:kor99}
\end{figure}

The kinematic axis is significantly different from the P1--P2 axis (PA $42\degr$) 
as shown in Fig.~\ref{fig:vmap}, and  P1 is offset $0\farcs12$ from the kinematic axis.
The fact that {\em P1 is not aligned on the kinematic major-axis} must be taken into 
consideration for the interpretation of the high spatial resolution \hst\ kinematical data 
which have been taken close to the P1--P2 axis.

The velocity dispersion map of the (best resolved) M2 datacube peaks at 270~\kms\  at 
$0\farcs3 \pm 0\farcs1$ on the anti-P1 side.
The peak is extended, and surrounded by a halo which itself extends above 
the nearly constant bulge velocity dispersion of $\sim 150$ \kms. 
The same structure is found in the M8 datacube, but with a lower contrast.
The dispersion is clearly asymmetric with respect to the UV peak, with a difference of 35~\kms\  at 
$\pm 1\farcs2$ along the kinematic axis. 
Note that the offset of the velocity dispersion peak is also present
in the  KB99 data (see their Fig.~4). 
The small difference is simply due to the slightly lower resolution of the \sis\ data 
(including the smoothing onto the $0\farcs35$ arcsec slit width). 
Such an offset was also present in the {\tt TIGER} data (Bacon \etal \cite{Bac94}), 
but with a larger value ($0\farcs7$), due to the lower spatial 
resolution ($0\farcs9$ FWHM).

\subsection{Bulge-subtracted velocity maps}
\label{sec:bulgesub}

We have used two different models for the (unknown) velocity dispersion of the bulge:
a constant value of 150~\kms, or the dispersion predicted by a simple
Jeans model using the combined gravitational potential of the nucleus
and the bulge (as in Bacon \etal\ \cite{Bac94}). The two resulting
kinematical profiles only show minor differences, the bulge-subtracted velocity
and dispersion for the ''constant dispersion'' model being slightly larger:
for clarity, we will only deal with the latter
(note that these profiles are consistent with the bulge-subtracted 
kinematics of KB99).

After the subtraction, the asymmetry in the velocity field is now clear, 
with the anti-P1 side having a slightly higher velocity amplitude
with a local minimum at $-250$~\kms\  (at $\sim 0\farcs75$
from the centre; fig.~\ref{fig:subulge}). The velocity profile on the P1 side
is nearly flat with a value of $\sim 180$~\kms.
The dispersion now peaks at 329~\kms\  at $-0\farcs1$ from the centre. 
We also confirms that the nucleus is cold (KB99), with a value of $108 \pm 10$~\kms\ 
at $1\farcs2$ along the kinematical major-axis on the P1 side.
\begin{figure}
\centering
\resizebox{8cm}{!}{\includegraphics[clip]{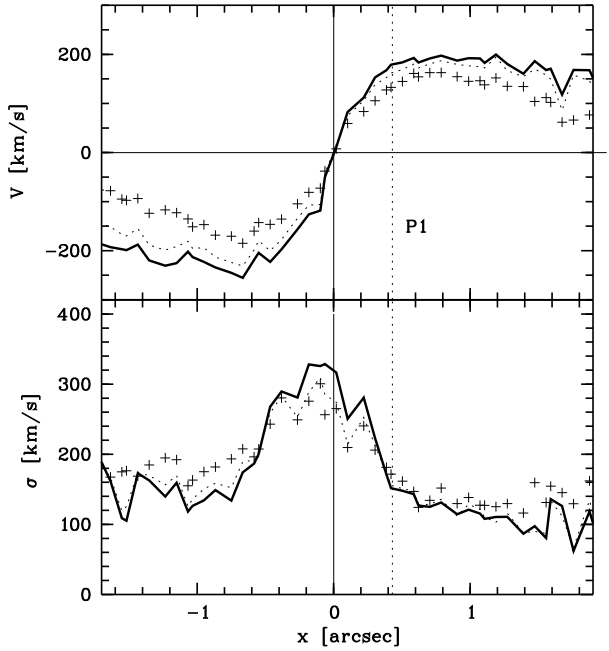}}
\caption{Bulge-subtracted \oasis\ kinematics along the kinematic 
major-axis of the nucleus compared with the original \oasis\ M8 profiles (crosses).
The solid line shows the resulting kinematics when using a constant
dispersion of 150~\kms\  for the bulge, and the dotted line when
using the dispersion predicted by a simple Jeans model.}
\label{fig:subulge}
\end{figure}

\subsection{\stis\ and \foc\ kinematical profiles}

The \stis\ velocity and dispersion profiles\footnote{\stis\ data are available in ascii/fits form
	at {\tt http://www-obs.univ-lyon1.fr/$\sim$tigerteam/bacon01.html}} 
are presented in Fig.~\ref{fig:stis}.
There is a clear asymmetry in the velocity curve, with
local turnover values of $197 \pm 5$ and $-292 \pm 20$~\kms\  at radii of $+0\farcs42$
and $-0\farcs235$ respectively. The maximum velocity on the P1 side
is $201\pm5$~\kms\  at $0\farcs63$. The dispersion is maximum on the anti-P1 side at a radius of
$-0\farcs235$ with a value of $321\pm33$~\kms\  (and a value of $313\pm36$~\kms\  at $-0\farcs18$). 
This is to be compared with
the offset of $0\farcs15$--$0\farcs20$ found\footnote{This includes
the spatial offset of $0\farcs031$ discussed in Sect.~\ref{sec:centre}.}	
by KB99 and $\sim 0\farcs2$ for the slit profile reconstructed 
from the \oasis\ data. The \stis\ velocity profile
crosses the $V=0$ line at $+0\farcs09$ (P1 side). At $[0,0]$ we measure $V=-68\pm6$~\kms.
After subtraction from the bulge contribution, the central velocity gradient is slightly
steeper, with turnover values of $-355$~\kms\ and $221$~\kms\ at $-0\farcs235$ and $0\farcs47$
respectively. 
\begin{figure}
\centering
\resizebox{8.cm}{!}{\includegraphics[clip]{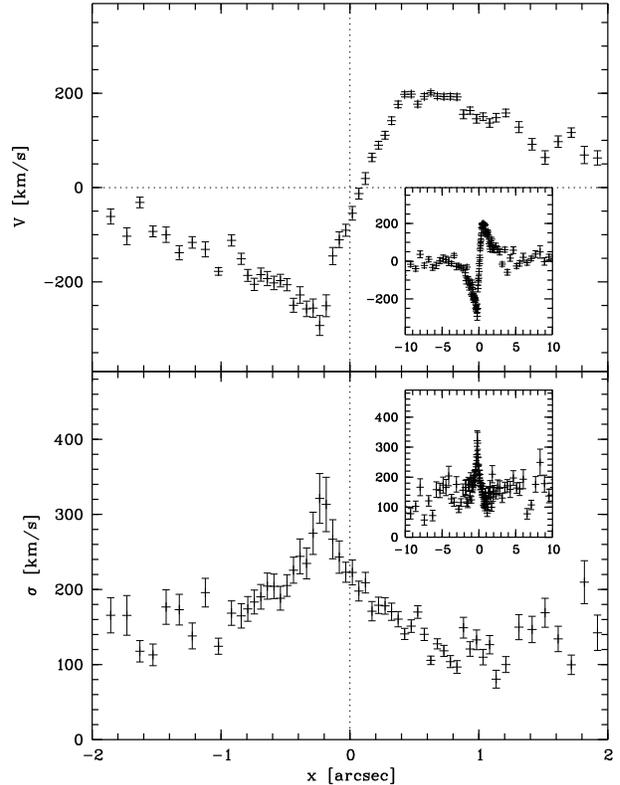}}
\caption{\stis\ kinematical profiles: velocity (top panels) and dispersion
(bottom) panels at a PA of 39$\degr$. The inserted panels present the \stis\ kinematical
profiles within a radius of 10\arcsec.}
\label{fig:stis}
\end{figure}
\begin{figure}
\centering
\resizebox{7.5cm}{!}{\includegraphics[clip]{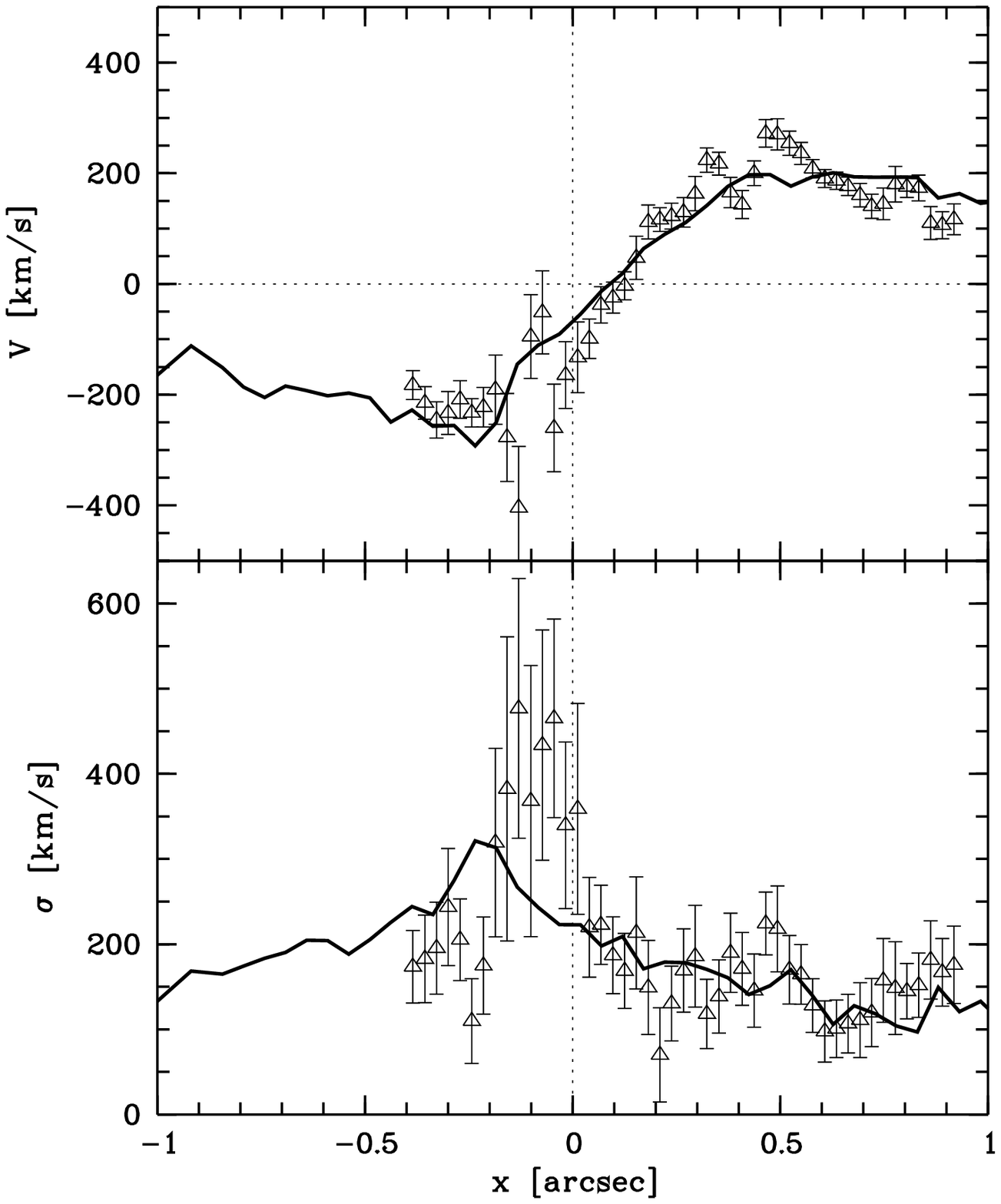}}
\caption{Comparison between \stis\ (solid line) and \foc\ kinematics (filled triangles).
A shift of $0\farcs025$ towards the anti-P1 side was applied, following
the analysis made in Sect.~\ref{sec:centre}. A positive
shift of $30$~\kms\ was applied to the \foc\ velocity profile (top panel, see text)}. 
The \stis\ data and the \foc\ data were 
taken at PAs of 39$\degr$ and 42$\degr$ respectively.
\label{fig:stisfoc}
\end{figure}

Comparing the original \foc\ (Sta+99) and
the \stis\ data, we find significant
discrepancies in both the velocity and dispersion profiles at the very centre
which seem difficult to attribute to differences in instrumental 
characteristics. The most surprising difference is the location of the dispersion peak
in the original \foc\ data which is nearly centred with an offset of only $0\farcs06$
from the UV peak (away from P1). By including the observed spatial shift of $0\farcs025$
mentioned in Sect.~\ref{sec:centre}, as well as a velocity shift of $30$~\kms,
the comparison looks reasonable (Fig.~\ref{fig:stisfoc}): this point is discussed
below (Sect.~\ref{sec:compstis}).

\subsection{\oasis\ compared with \hst\ kinematics}
\label{sec:compstis}
Although the \stis\ and \foc\ data have similar slit widths 
(0\farcs1 for \stis\ and 0\farcs063 for \foc)
and PAs ($39\degr$ for \stis\ and $42\degr$ for \foc), the \foc\ data (Sta+99)
have been obtained at much bluer wavelengths than \stis\ (0.45 versus 0.85 $\mu$m) 
and have a better spatial resolution (Fig.~\ref{fig:stisfoc}).
A detailed comparison between the \stis\ and \foc\ data sets thus requires
to take into account these specific characteristics. 
But this implies the convolution of the (yet unknown) 2D kinematics of 
\M at \foc\ spatial resolution. We also wish here to compare \hst\ and \oasis\
kinematical data: the two-dimensional coverage of the \oasis\ spectra
is in this context a powerful additional constraint.
The comparisons between the \oasis\ and \stis\ datasets are shown in Fig.~\ref{fig:stisoakb}.
\begin{figure}
\centering
\resizebox{8.cm}{!}{\includegraphics[clip]{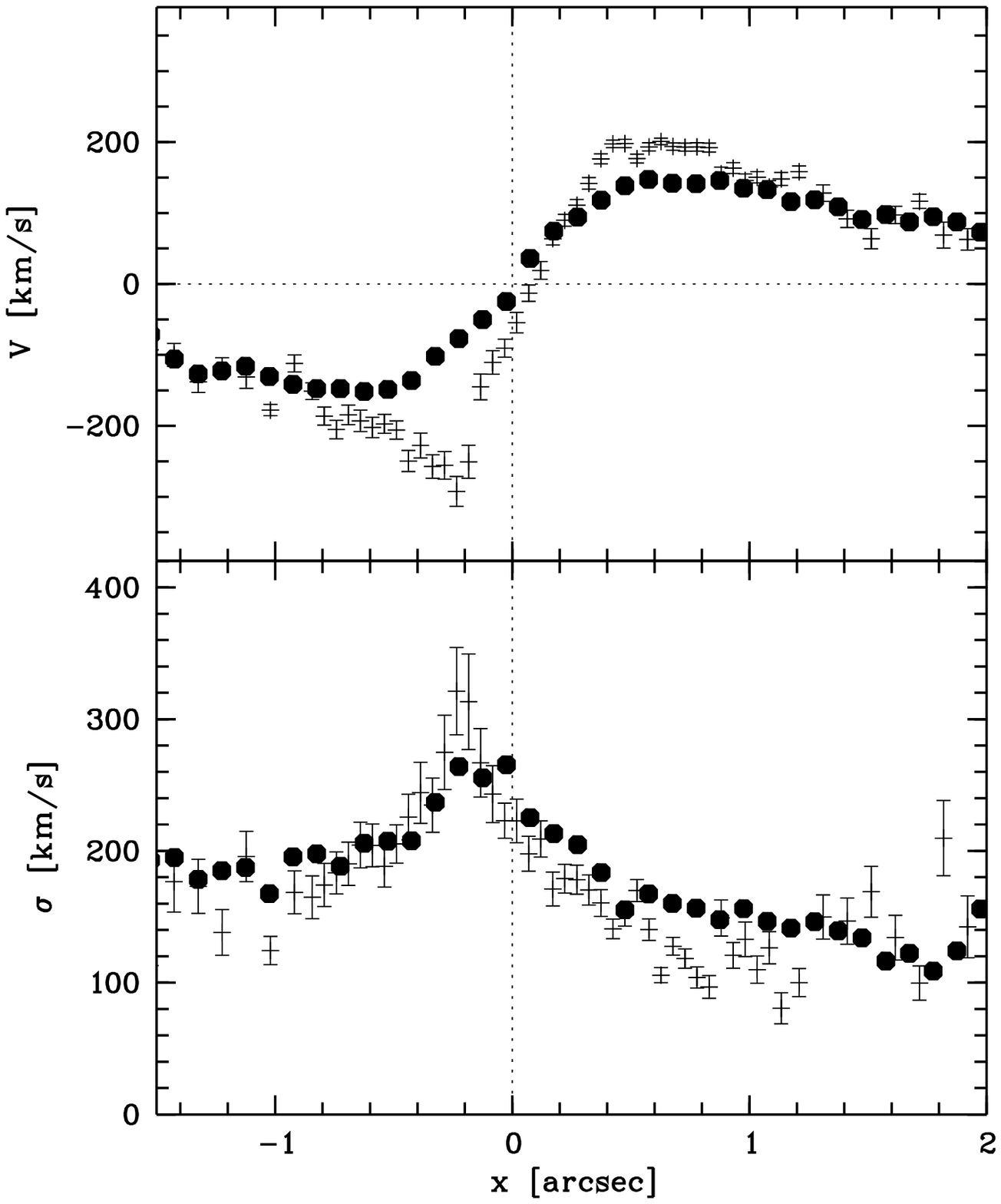}}
\caption{Comparison between the kinematics from \stis\ (crosses) and \oasis\ (filled circles).
The \oasis\ kinematics have been averaged over a $0\farcs2$ wide slit
(PA$=39\degr$).}
\label{fig:stisoakb}
\end{figure}

To attempt such a comparison,
we have built simple 2D parametric kinematical models. These models
are obviously ill-constrained, given that we have knowledge on only 
a tiny fraction (2.5\%) of the area covered by \oasis\ at \foc\ resolution,
and do not stand on any physical ground.
This {\em ad hoc} model is solely designed to check if
the three kinematical data sets are consistent with each other, i.e.
if we can find a reasonable model which simultaneously fits all the observed kinematics.
All convolutions are achieved taking into account the relevant pixel
integration. The surface brightness distributions were derived
from the deconvolved \wfpc\ images: we kept the F814W filter image
for both the \stis\ and \oasis\ data, and used a weighted sum
of the F300W and F555W images to roughly reproduce the surface brightness
profile of the \foc\ data presented by Sta+99.
In the cases of the \oasis\ and \stis\ data, we used an estimate
of the first two true velocity moments (the best gaussian 
velocity and velocity dispersion depending on the shape of
the line-of-sight velocity distribution). For the \foc\ data,
we had to use the original values as published by Sta+99.
Since Sta+99 mention that the kinematics derived from
the half blue and red parts of their spectra are indistinguishable, 
we assumed in the following that the UV peak does not
contribute to the kinematical profile. In other words, we 
assumed that the stars forming the UV peak do not
show (metallic) features strong enough to significantly contribute to
the observed kinematics: this hypothesis seems to be supported
by a preliminary analysis of the \stis/G430L data (Emsellem~\etal, in preparation).
\begin{figure*}
\centering
\resizebox{15cm}{!}{\includegraphics[clip]{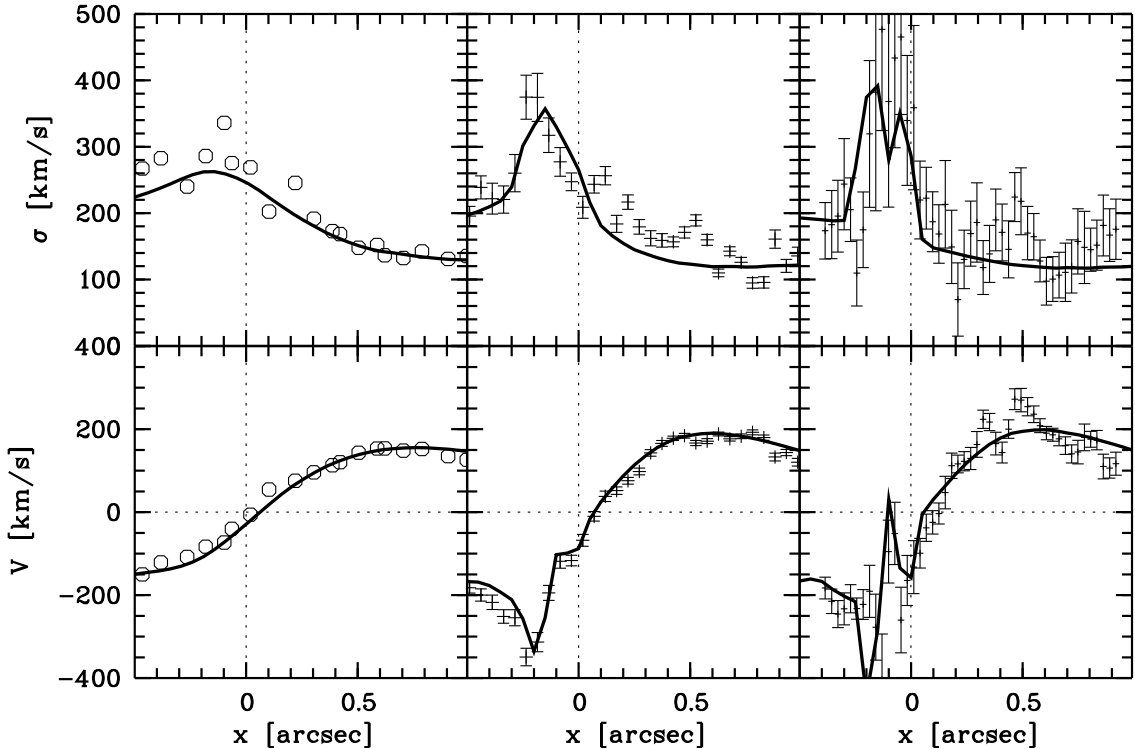}}
\caption{Comparison of \oasis\ (left), \stis\ (middle) and \foc\ (right)
kinematics within the central arcsec. Velocity and dispersion profiles
are shown in the bottom and top panels respectively. Original \foc\ data 
(right panels) have been shifted by $0\farcs025$ and $30$~\kms\  
(see text). The solid lines represent the model 
convolved and sampled according to the corresponding instrumental setup. 
}
\label{fig:mod2zoom}
\end{figure*}
\begin{figure}
\centering
\resizebox{8.cm}{!}{\includegraphics[clip]{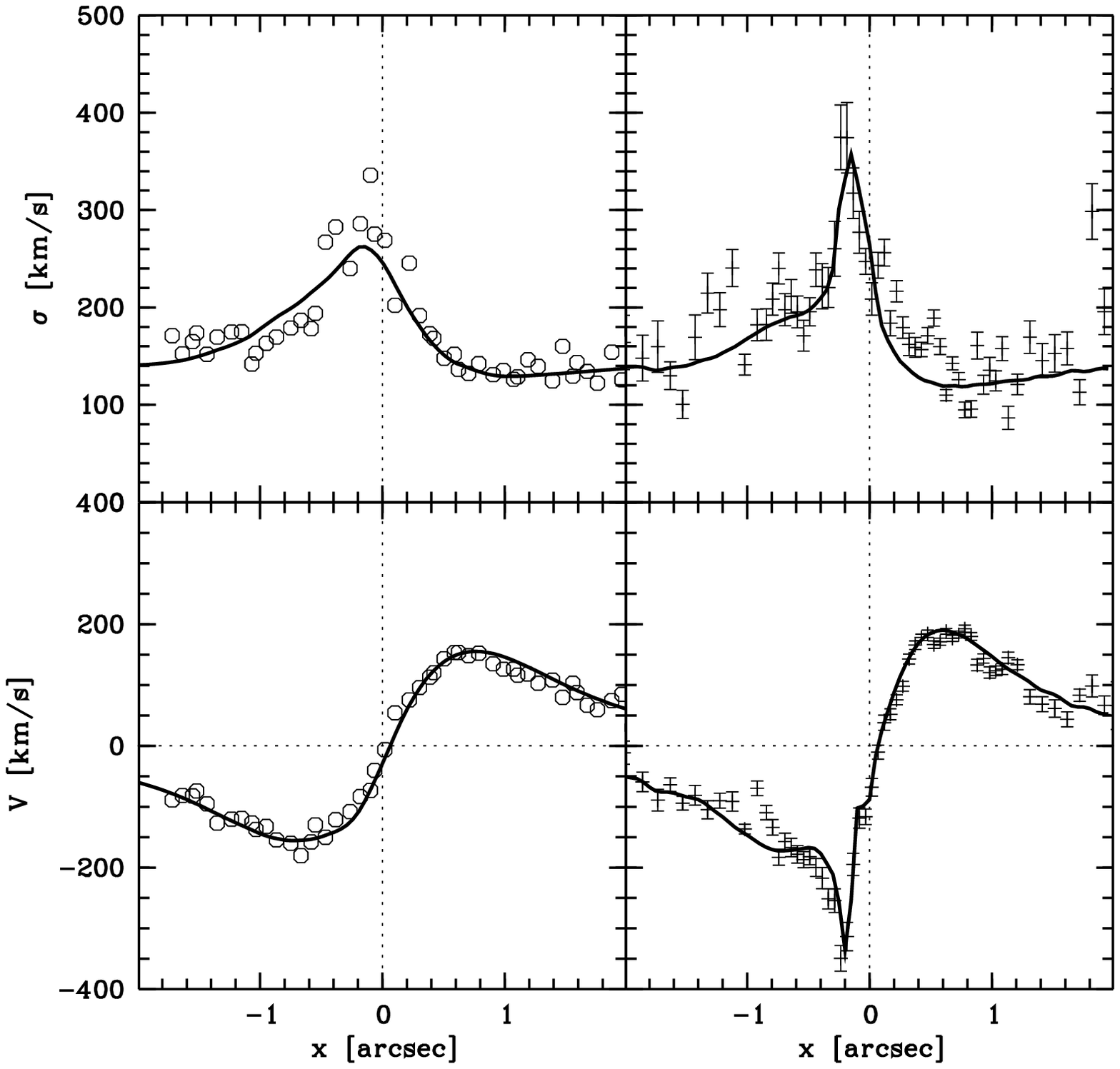}}
\caption{Same as in Fig.~\ref{fig:mod2zoom} but for the \oasis\ and \stis\ data
in the central 2\arcsec.
}
\label{fig:mod2}
\end{figure}

We start with a model based on velocity functions of a simple form 
$V(r) = K \cdot x \cdot (1 + r^2/r_e^2)^{-n}$ with $r^2 = x^2 + y^2/q^2$
($x$ and $y$ being the cartesian coordinates aligned with the kinematical
axis), and constant or 
gaussian functions for the velocity dispersion. The PA of the kinematical major-axis 
was set to 56.4$\degr$ as measured from the \oasis\ data. The observed asymmetries in the 
kinematical profiles were obtained by allowing both spatial
and velocity offsets in the parametrized functions.
We fixed the dispersion of the bulge and the nucleus to 150~\kms\  and 108~\kms\ 
respectively, adding (quadratically) a gaussian of $1\farcs2$ FWHM with a maximum
of 140~\kms\ located at $0\farcs8$ on the P1 side to reproduce the 
low-frequency spatial asymmetry seen in the \oasis\ dispersion map (Fig.~\ref{fig:vmap}).
The velocity and the non-centred second order moment ($V^2 + \sigma^2$) 
are then convolved, after proper luminosity weighting. 
The input parameters were tuned 
until a satisfactory result was obtained.

The best overall fit was obtained by adding high velocity
components in the central $0\farcs3$ aligned with
the kinematic major-axis. The dispersion peak then
simply results from the effect of velocity broadening 
as shown in Fig.~\ref{fig:mod2}. We cannot distinguish
between a velocity broadening effect from true velocity dispersion of components
that are spatially unresolved. This model is anyway the simplest 
we found which provides a reasonable fit to the \stis\ and \oasis\ data.
It includes maximum amplitude velocities of $\sim +235$ and $\sim -265$~\kms\ 
(outside the central $0\farcs3$).
We also confirm the spatial and velocity offsets 
in the {\em original} \foc\ data with respect to the UV peak
and the zero velocity as measured by KB99, respectively: by 
shifting the \foc\ kinematics by $0\farcs025$, consistently with 
the analysis of Sect.~\ref{sec:centre}, and adding
30~\kms\ to their systemic velocity, we can indeed reconcile
the model with the \foc\ data (Fig.~\ref{fig:mod2zoom}).
Note that this velocity shift is consistent with the maximum possible 
systematic error of $50$~\kms\  quoted by Sta+99.
The high velocity component of the model is an attempt
to reproduce the abrupt jump seen in the \foc\ rotation curve at a radius $\sim -0\farcs1$,
near the location of the dispersion peak.
This feature should however not be overinterpreted, although we believe that
the central dispersion peak can indeed be explained via velocity broadening
with the contribution of fast moving stars within $0\farcs3$ of the UV peak
on the anti-P1 side. We do not find any significant discrepancy
in the dispersion profiles contrarily with what was advocated by Sta+99:
as mentioned earlier (Sect.~\ref{sec:centre}), they uncorrectly 
shifted KB99's data to manage a consistent fit (their Fig.~6). 

We cannot make any definite statement about the detailed kinematics in the central
$0\farcs3$, as there is too much freedom in the model parameters. This is
mainly due to the fact that \foc\ and \stis\
provide only one-dimensional profiles, unlike \oasis\ data which are two-dimensional,
but have a too low spatial resolution to constrain the dynamics at that scale.
We can however still make a few relatively safe statements, which will
be discussed further in Sect.~\ref{sec:conclusion}:
\begin{itemize}
\item The observed kinematics is definitely not
consistent with a dispersion peak at the location of the UV peak (our $[0,0]$).
\item The dispersion peak and high velocities seem to be spatially 
associated with P2 (as seen in the $I$ band) which is offset from the UV
peak on the anti-P1 side as shown in Fig.~\ref{fig:orient}. 
\item The \oasis, \stis\ and \foc\ kinematical profiles are consistent with each other,
although a two-dimensional coverage of the central $0\farcs3$ is required
to properly address the kinematics there.
\end{itemize}

\section {Numerical Simulations}
\label{sec:simu}

We now interpret the observations with the help
of N-body modeling. Among the various hypothesis
that have been advanced to explain the M~31 double nucleus (with luminosity peak P1 shifted
by $\sim$ 1.8 pc from the kinematical centre, almost coinciding with the
UV peak), the most natural would
be an $m=1$ wave in a rather cold and thin disk orbiting the SBH, 
located at the centre of the UV peak.
The disk has to be cold, to be unstable to non-axisymmetric waves, and this implies 
a small thickness, in the almost spherical potential provided by the central SBH.
However it is still unclear whether $m=1$ perturbations, accompanied
by a displacement of the gravity centre, will be unstable, and develop
spontaneously in the physical conditions corresponding to the M~31 nucleus
 (Heemskerk et al. \cite{Hee92}, Lovelace et al \cite{Lov99}).

An alternative solution is that the disk is initially perturbed by an external force (either 
a globular cluster or Giant Molecular Cloud passing by), and the response is long-lived
with a time-scale comparable to that of such external perturbations. We
have simulated this possibility, and indeed found modes of $m=1$ oscillations
that maintained during 70 Myr, with an almost constant pattern speed. This peculiar
feature is due to the very low precession rate in an almost Keplerian disk, near a SBH.
The asymmetry of the density, and the radial variation of eccentricity of the orbits,
generate local variation of the effective precession rate, and most of
the stars are dragged into a mode of slow, positive pattern speed.
In this paper, we propose that this mechanism is at the origin of the
M~31 $m=1$ perturbation (or ``double-nucleus'' morphology), and illustrate
this with N-body simulations with asymmetric initial conditions.
We consider in a future work the possibility that the
$m=1$ instability develop spontaneously from axisymmetric initial conditions
(Combes \& Emsellem \cite{CE01}).

\subsection {N-body Methods and Diagnostics}

We performed essentially 2D simulations, since our interpretation is that 
the nuclear disk of M~31 is cold and thin (axis ratio of the
order 0.1).  The apparent axis ratio must then be mostly 
due to an inclination of $i \sim 55^\circ$ (see Sect.~\ref{sec:bestfit}), less edge-on than 
the large-scale M~31 disk ($i = 77^\circ$). However we also checked the results with 3D experiments. 
 The gravity is solved via fast Fourier Transforms, on a useful 2D grid of
$256\times256$ ($512\times512$ to suppress Fourier images). In 3D, the N-body code is
a FFT scheme, which uses the James  (\cite{Jam77}) method to avoid the influence
of Fourier images. The grid is then $128\times128\times64$. The size of the simulated box in 
the disk plane is 20~pc, corresponding to a cell size of  0.078~pc (and twice that in 3D).
This size is also the softening length of the Newtonian gravity. The M 31 nuclear stellar
disk is represented in 2D by 99000 particles  and 152384 in 3D. Two rigid
potentials are added, representing the bulge and the supermassive black hole. 
The time step is 10$^{-4}$ Myr. 

Particle plots are made in face-on and M~31-sky projections, together with the velocity
field, the density, velocity and dispersion profiles along the major axis, to compare
with the observed quantities (with and without bulge addition). The Fourier analysis
of the density and potential are made regularly as a function of time and radius. 
If the potential  is decomposed as
$\Phi$(r,$\theta$) = $\Phi_0$(r) + $\Phi_m$(r) cos (m $\theta$ - $\phi_m$),
we define the intensities of the various components by their maximal contribution to the
tangential force, normalised by the radial force $ F_r = -\partial \Phi_0 /  \partial$r, through
$$
S_m = m \Phi_m / r | F_r |
$$

In some simulations the SBH is allowed to move, with respect to the
fixed background bulge. However, this introduces additional oscillations, of small
amplitude, which are not well computed with the low spatial resolution in 3D. 
Therefore in most cases, the SBH was fixed at the centre of the bulge.
These oscillations are of different (higher)  frequencies that the phenomenon 
we are studying here, and are only superposed to it. They may be related
to the nuclear oscillations studied by  Taga \& Iye (\cite{Tag98}) and
Miller \& Smith (\cite{Mil92}). 

Over all run periods of typically 10-100 Myr, the Fourier analysis of the disk potential
was done as a function of radius, and stored every 0.03 Myr. The result $F(r,t)$ is then
Fourier transformed in the time dimension, in order to get the power as a
function of pattern speed, $\tilde{F}(r, \Omega)$. With respect to the frequency of the $m$ 
perturbation, the pattern speed is $\Omega = \omega/m$.

\subsection{Galaxy Model and Initial Conditions}
\label{galmod}

The massive black hole potential was softened to a few cell sizes to
avoid prohibitively small values of the time step
in the centre. It is represented by a Plummer shape potential,
$$
\Phi_{BH}(r) = - {{G M_{\rm BH} }\over {\sqrt{r^2 +r_{bh}^2}}}
$$
with $r_{bh}=0.07$~pc. The mass of the SBH  $M_{\rm BH}$ 
was varied between 3.5 and 10~10$^{7}$ M$_\odot$
to test the role of the self-gravity of the disk. The fixed analytical potential
of the bulge is composed of 4 Plummer
functions, determined from the MGE method (see Emsellem \& Combes \cite{EC97}).
 Their corresponding masses and radii are displayed in Table \ref{param}.
 The total mass of the bulge inside 9~pc is 0.87~10$^{7}$ M$_\odot$. 
The nuclear stellar disk is initially a Kuzmin-Toomre disk of surface density
$$
\Sigma(r) = \Sigma_0 ( 1 +r^2/d^2 )^{-3/2}
$$
truncated at 9~pc, with a mass  of $M_d = 1.7\, 10^{7}$~M$_\odot$, and 
characteristic radius of $d = 3.5$~pc.
It is initially quite cold, with a Toomre Q parameter of 0.3. The mass fraction of the
disk inside 9~pc was varied between 20 and 40\%.

Since the potential of the black hole is significantly smoothed below
a radius equal to 2.5 times the softening length, here $2.5 \times 0.07$~pc$= 0.175$~pc,
we have avoided putting particles in this region, so as not to introduce
artificial dynamics. The initial surface density was then equal to the
difference between two Toomre disks of the same central surface
density, but different characteristic lengths. This was done to have
an analytic density-potential pair with a smooth boundary for
the central hole. Up to 12\% of the
central disk mass was thus removed and redistributed on the
remaining outer parts. The exact value of the scale-length of
the removed disk, between 0.5 and 1.2~pc, was of no importance
on the final results.
 It is somewhat larger than the required 2.5 softening lengths,
since the initial orbits near the black hole are eccentric, with
a small pericentre. 
The potential inside this radius is in any case dominated by the
keplerian law of the black hole, so that the suppression of the
central stellar disk does not affect the rotation curve.
The rotation curve obtained is  plotted in Fig.~\ref{omeg}.
\begin{table}[h]
\caption[ ]{Parameters of the mass model}
\begin{flushleft}
\begin{tabular}{ccc}  \hline
Component   & Size   &  Mass$^*$    \\
          &       pc & 10$^{7}$ M$_\odot$              \\
\hline
SBH         &  0.07    &  7         \\
Disk        &  3.5    &  1.7      \\
Bulge-1   &  700  &  0.01     \\
Bulge-2   &  160  &  0.07     \\
Bulge-3   &  50    &  0.2       \\
Bulge-4   &  15    &  0.6       \\
\hline 
\end{tabular}
\end{flushleft}
$^*$ mass inside 10 pc radius\\
\label{param}
\end{table}

The initial $m=1$ perturbation was produced in several ways: either
the SBH position was shifted by an arbitrary value from the centre of the potential,
or its velocity was perturbed and given a high value. 
In both cases, a long-lived $m=1$ pattern was obtained.
 However, the initial conditions were too far from equilibrium, and the violent relaxation
led too many particles to escape.  A softer way to perturb the disk is to launch particles in
eccentric orbits from the start, keeping the SBH fixed at one of the foci
of the elliptical orbits. Once a profile of eccentricity is chosen, each particle is moved
along the corresponding ellipse at a random position with probability inversely proportional
to its linear velocity at this point (probability density along the ellipse proportional to
$1 - e \cos{(\xi)}$, $\xi$ being the true anomaly). About 20 models were run, to investigate
the nature of the modes, and the influence of initial conditions, the mass fraction
of the disk, etc. The best fit for M~31 is described here, and compared with observations.
\begin{figure}
\centering
\resizebox{8.2cm}{!}{\includegraphics[clip=true,angle=-90]{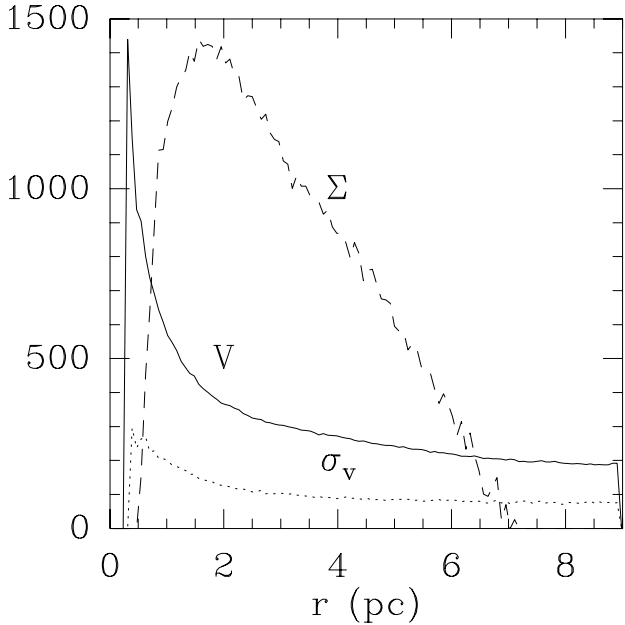}}
\caption[omeg.ps]{ The observed orbital velocity $V$ (full line, in km/s), 
log of surface density in arbitrary units ($\Sigma$, dash), and radial velocity dispersion  
($\sigma_v$, dot-dash), at the epoch of 54 Myr,
for the galaxy model with $M_{\rm BH} = 7\, 10^7$ M$_\odot$.}
\label{omeg}
\end{figure}

\subsection{ Results}
\label{sec:bestfit}

The best fit model for M~31 was obtained through the initial distribution
of eccentricity
$$ 
e(a) = 0.6 \sqrt{( 1 - (a/4\; \mathrm{pc})^2)}
$$
and $ e(a) = 0$ for $ a > 4 pc$ ($a$ being the semi-major axis).
The major axis of the orbits are initially aligned,
but they quickly re-arrange, so that the pericentre of orbits in the inner and
outer parts are in phase opposition. The face-on view of the nuclear
disk is displayed in Fig.~\ref{cont_m19fpc}. There is a clear density 
accumulation on one side of the BH. The strength
of the $m=1$ and $m=2$ Fourier components of the potential are
plotted in Fig.~\ref{fig:p1cet_m19fpc}, together with the position of the
stellar disk gravity centre, as a function of time. At the beginning, until
15 Myr, there is a settling phase, where the $m=1$ mode has a
trailing, than clearly leading arm. The stellar 
 gravity centre precesses in the positive direction, and also
the pattern speed in the power spectrum is positive (see Fig.~\ref{fig:pow_m19fpc}).
Then, the motion of the gravity centre becomes more regular,
slows down slightly, with a period of 2 Myr,
corresponding to a pattern speed of the $m=1$ perturbation of
 $\Omega_p \sim  3$~km/s/pc (the pattern speed is equal to the orbital 
frequency of the gravity centre). Note that the $m=2$ perturbation is just the harmonic
of the $m=1$, and has the same pattern speed.

These values of pattern speed are about 10 times smaller than the value
derived by Sambhus \& Sridhar (2000) using the Tremaine-Weinberg method
and M~31 FOC data. Note however that the latter method is rather uncertain
when applied to one-dimensional data.
\begin{figure}
\centering
\resizebox{6.2cm}{!}{ \includegraphics[clip=true]{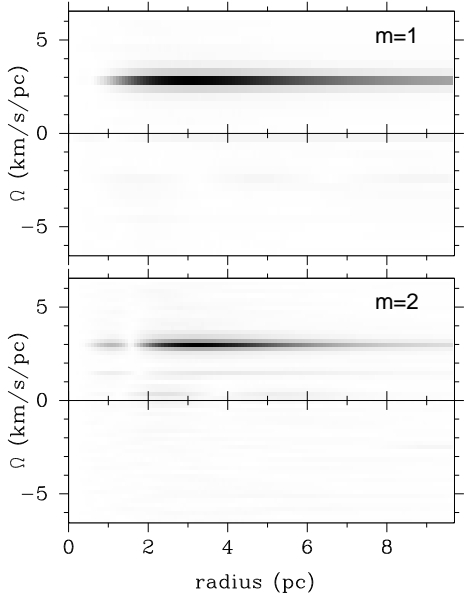}}
\caption[pow_m19fpc.ps]{Pattern speed as a function of radius, in units of km/s/pc, for
the $m=1$ mode (top) and the $m=2$ mode (middle) between the epochs 28.8 and 43.2 Myr
($M_{\rm BH} = 7\, 10^7$ M$_\odot$).
The pattern speed slows down slightly with time, which thickens the lines (from 3 to 2.5
 km/s/pc in 60 Myr).}
\label{fig:pow_m19fpc}
\end{figure}

The distribution of eccentricity is quite similar all over the run, and 
typically represented by Fig.~\ref{eda_m19fpc}. It is always decreasing
from $ a = 1$~pc to 6~pc, with a large scatter. The eccentricity
is maintained at large values, and the apocentres are aligned to give the
density accumulation that will give rise to P1 in M~31 (see the position
of P1 in the middle and bottom of 
Fig.~\ref{eda_m19fpc}). From 5 to 10~pc, the apocentres change side,
and there is a secondary density accumulation, in phase opposition
with S1, of much smaller amplitude. The accumulation is also at
the apocentre of these outer particles. The density accumulation
is located very close to the maximum eccentricity gradient. 
As noted by Statler (\cite{Sta99a}),
this is a configuration that allows the right impulses on eccentric orbit
to equalize the precessing rates. If an orbit grazes from the inside 
a density accumulation, the experienced outward pull  can accelerate
in the positive sense its precession, if it is at apocentre. If it is
at pericentre, its precession will be slowed down. The nearly axisymmetric
(epicyclic approximation) precessing rate is only slightly negative, 
and this small impulse is sufficient to align the orbits, and
make them precess in the positive sense.
The orbits should be at their apocentre
inside the density peak, and at their pericentre inside, for the
precessing corrections to go towards equalization. This explains
the orbit orientation observed, and the fact that the two density
accumulations are in phase opposition.
\begin{figure}
\centering
\resizebox{8.2cm}{!}{\includegraphics[clip=true,angle=-90]{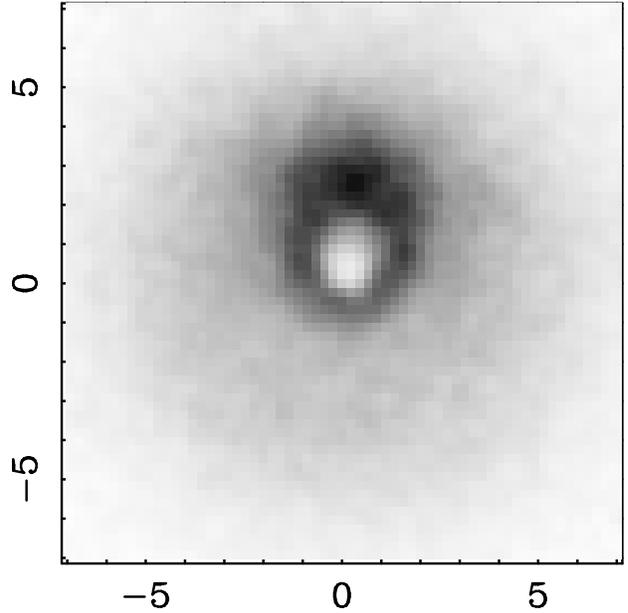}}
\caption[cont_m19fpc.ps]{Face-on surface density of the nuclear stellar disk, at epoch
28.8 Myr ($M_{\rm BH} = 7\, 10^7$ M$_\odot$). 
The scale is in pc. The hole in the centre is introduced on purpose in the
initial conditions, to avoid artificial effects due to the black hole softening and limited
spatial resolution (see Sect.~\ref{galmod}).}
\label{cont_m19fpc}
\end{figure}
\begin{figure}
\centering
\resizebox{5.2cm}{!}{\includegraphics[clip=true,angle=-90]{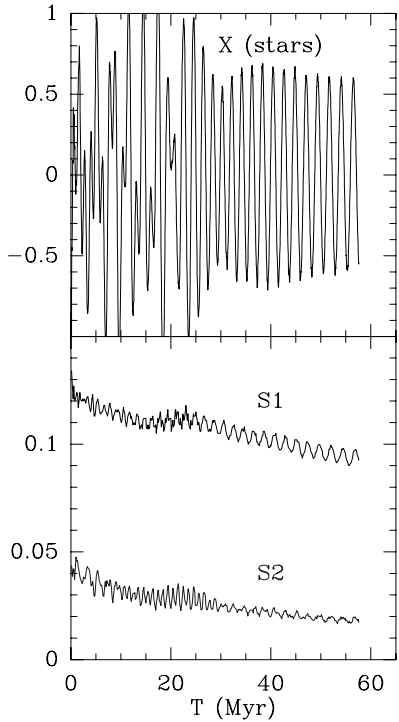}}
\caption[p1cet_m19fpc.ps]{{\bf top:} Coordinates (X: full line) of the centre of gravity of
the stellar disk, as a function of time (model with $M_{\rm BH} = 7\, 10^7$ M$_\odot$). 
Note that the period is slightly increasing with time, i.e. the pattern speed slows down.
{\bf bottom:} Intensity of the $m=1$ (S1, solid line) and 
$m=2$ (S2, light line) Fourier components of
the potential (more exactly the corresp. components of the tangential force normalised by
the radial force).} 
\label{fig:p1cet_m19fpc}
\end{figure}

The morphology and kinematics of the simulated nuclear disk at 
epoch $T = 10$~Myr is compared to the M~31 observations in 
Figs~\ref{fig:compnbody} and \ref{fig:compvnbody}. With an inclination on
the plane of the sky of $i = 55^\circ \pm 5$, and a position angle of the density
maximum such that P1 is not on the line of nodes, the comparison is
reasonably satisfying.  The constraint on the inclination angle is obviously
model dependent. However, the disk needs to be thin for
non-axisymmetric wave to develop. For inclinations $i > 60^\circ$,
it will thus be hard to reconcile our proposed model
with the observations. There is of course a deep hole in the centre of 
the simulated image, since we have avoided placing particles here
(see Sect.~\ref{galmod}). The limited spatial resolution imposes
a softening of the BH to 0.1~pc scale, and prevents following the stellar
dynamics there. For Figs~\ref{fig:compnbody} and \ref{fig:compvnbody}, we have artificially filled in this hole
by adding a central axisymmetric disk of particles: they do not participate
in the $m=1$ mode but their velocities are of course chosen according to the 
background potential.
\begin{figure}
\centering
\resizebox{8.2cm}{!}{ \includegraphics[clip=true]{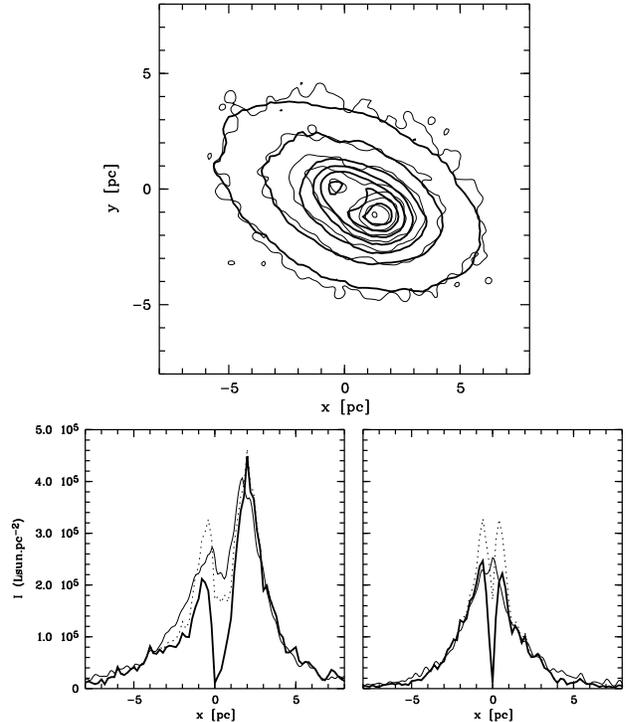}}
\caption[fig_comp_nbody.eps]{Comparison between the bulge-subtracted 
	WFPC2/HST $I$ band surface brigtness of the nucleus of M~31 (thin solid line) 
	and an N-body simulation of an $m=1$ mode where the disk
	was 40\% of the central mass (with and without the addition
	of particles in the central hole as mentioned in the text, 
	represented by dotted and thick solid lines respectively). The inclination
	of the disk was chosen to be $i = 55^\circ$.
	Top: isophotes in the central 8~pc. Bottom: cuts along (left) and perpendicular (right) to
	the P1--P2 axis.
}
\label{fig:compnbody}
\end{figure}
\begin{figure}
\centering
\resizebox{8.2cm}{!}{ \includegraphics[clip=true]{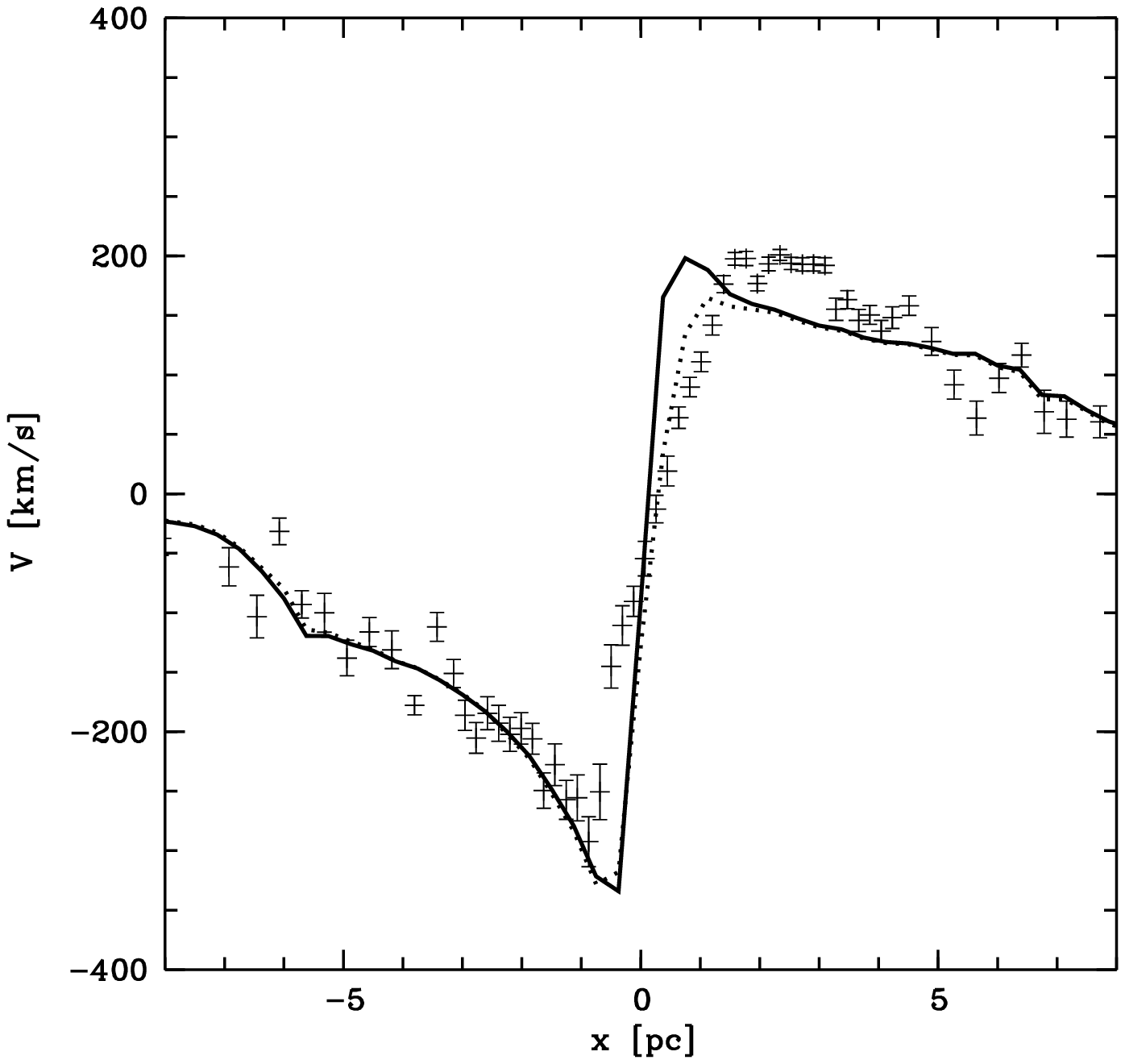}}
\caption[fig_comp_nbody.eps]{Comparison between the STIS/HST velocity profile (crosses
	with error bars) and the corresponding profiles from an N-body simulation of an $m=1$ mode
	where the disk was 40\% of the central mass (with and without the addition of 
	particles in the central hole; dotted and thick solid lines respectively).}
\label{fig:compvnbody}
\end{figure}

Although the presented model has many limitations, it reproduces the most
important observed features, specifically the asymmetries in the 
surface brightness between P1 and its phase opposite (Fig.~\ref{fig:compnbody}) as well as in the 
velocity profile (Fig.~\ref{fig:compvnbody}). 
A thorough comparison, e.g. including the central velocity dispersion profile,
should wait for simulations including a more realistic treatment of the 
particles near the SBH.
We can however already add an important comment concerning stars in the central
parsec. We expect the stars in this region to participate to the $m=1$ mode, as observed
in self-consistent simulations where we initially did not remove the central particles.
In that case the sign of the eccentricity was reversed in the central parsec, creating
an offset light peak on the anti-P1 side. This configuration was already suggested
by Statler~(\cite{Sta99a}) who studied the sequence of closed orbits in a wide precessing disk.
It is consistent with the observed offsets of P2 ($I$ Band) surface brightness, and of the
dispersion peak on the anti-P1 side (see Sect.~\ref{sec:compstis} and Fig.~\ref{fig:orient}). 
The UV peak would thus still mark the location of the SBH, but it would not contribute to the measured
kinematics. 

\subsection{ Discussion}

The $m=1$ modes of a self-gravitating disk, dominated by a central point
mass, have not been fully investigated. 
Jacobs \& Sellwood (\cite{Jac99}) have reported the presence of a slowly 
decaying $m=1$ mode in annular disks around a slightly softened
point mass, but only for disk masses less than 10\% of the central mass 
concentration.
It is expected that there is a threshold
mass of the central BH, above which eccentric modes could be 
sustained by gravity. Indeed, in the extreme cases, where the disk mass
is negligible, the orbits are Keplerian, and their precessing rate is exactly
zero ($\Omega = \kappa$). If the apsides are aligned at a given time,
they will stay so in a $\Omega_p =0$ mode. The self-gravity of the disk
makes $\kappa > \Omega$, and the orbits differentially precess
at a rate $\Omega - \kappa < 0$ . However, if the disk self-gravity is not
large, a small density perturbation could be sufficient to counteract the 
small differential precession. Goldreich \& Tremaine  (\cite{Gol79}) 
showed that in the case of Uranian rings, the self-gravity could
provide the slight impulse to equalize the precessing rates, and 
align the apsides. Levine \& Sparke (\cite{Lev98}) proposed that lopsidedness
could survive if the disk is orbiting in an extended dark halo,
provided it remains in the region of constant density of the halo
(or constant $\Omega$, but this does not apply to M~31).
Lovelace \etal\ (\cite{Lov99}) found through linear analysis slowly growing modes, 
in the outer parts of a disk orbiting a central point mass. The pattern 
speed could be either positive or negative. Taga \& Iye (\cite{Tag98})
found by N-body simulations that a massive central body
can undertake long-lasting oscillations, but only when its mass is lower
than 10\% of the disk mass (which again does not apply to M~31). 

Here we propose that a natural $m=1$ mode can explain the M~31 
eccentric nuclear disk. We have found that for a disk mass
accounting for $\sim$  20 -- 40\% of the total central mass,
self-gravity is sufficient
to counteract the differential precession of the disk.
An external perturbation
can excite this mode, and it is then long-lasting, over
100~Myr, or 3000 rotation periods. The interval between two
such external perturbations (either passage of a globular cluster,
or a molecular cloud) in M~31 is of the same order of magnitude,
so that the external perturbations are an attractive mechanism.
 Each episode of $m=1$ waves will heat the disk somewhat,
but the instability is not too sensitive to the initial radial
velocity dispersion. Over several $10^8$~yr periods, the nuclear
disk could be replenished by fresh gas from the large-scale M~31
disk and subsequent star formation.
\begin{figure}
\centering
\resizebox{8.2cm}{!}{
\includegraphics[clip=true]{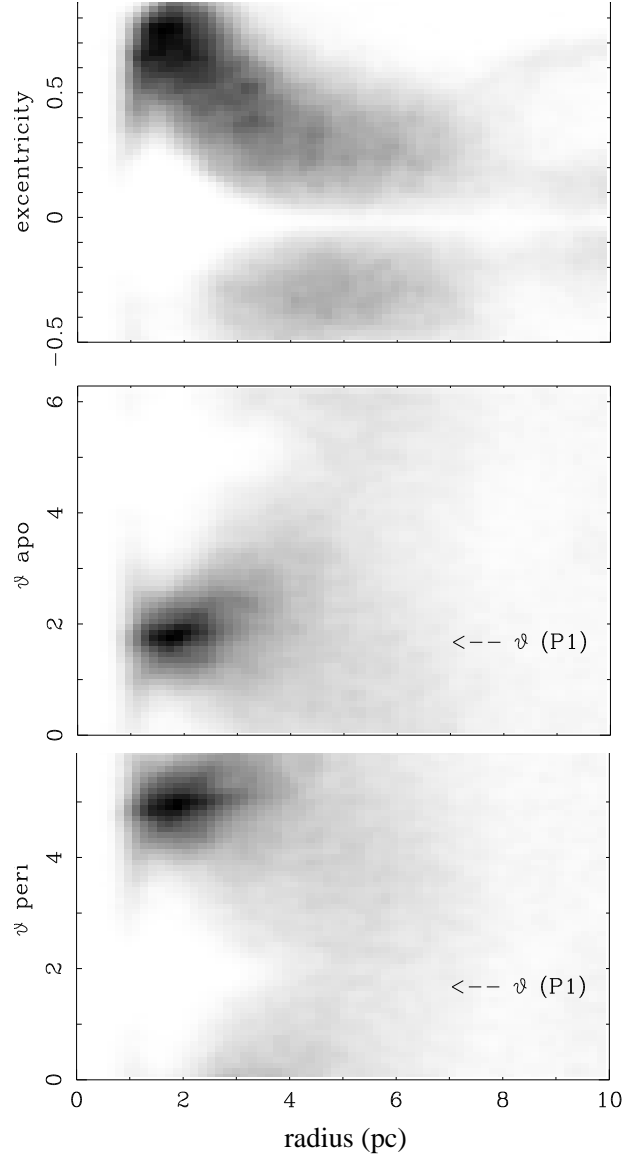}}
\caption[eda_m19fpc.ps]{Eccentricity $e(a)$ of particles as a function of 
their major axis $a$ {\bf (top)};  $e(a)$ is counted positive when the apocentre
is in the region of the P1 accumulation, negative otherwise.
Polar angle (from the BH) $\theta_{apo}$ of the apocentre of orbits {\bf (middle)}, and
 $\theta_{peri}$ of their pericentre {\bf (bottom)}, at epoch 28.8 Myr. 
The position of the maximum
density (corresponding to the P1 component of M~31) is indicated.}
\label{eda_m19fpc}
\end{figure}

\subsection{Comparison with linear WKB predictions}

In the potential of a point mass, orbits are exactly keplerian,
the precession rate $\Omega - \kappa$  of
eccentric orbits is zero. The presence of a small disk of mass $M_d$, lighter than the 
central point mass $M_{\rm BH}$, makes this precession rate negative, with amplitude varying as 
$\frac{M_d}{\sqrt{M_{\rm BH}}}$. If self-gravity has a large enough role, and in 
particular, if the disk is cold enough and its Jeans length smaller than the disk radius,
$m=1$ density waves can propagate; their dispersion relation has been studied
in the WKB approximation (Lee \& Goodman~\cite{Lee99}, Tremaine~\cite{Tre01}). In the linear
approximation, the pattern speed for the wavelength $\lambda = 2 \pi / k$ is in
first approximation for $\frac{M_d}{M_{\rm BH}} << 1$:
$$
\Omega_p = \Omega - \kappa + {{\pi G \Sigma_d |k|}\over{\Omega}} 
{\cal F}({{k^2c^2}\over{\Omega^2}})
$$
where $\Sigma_d$ is the surface density of the disk, and $\cal F$ the usual reduction
factor that takes into account the velocity dispersion $c$ of the stellar disk, and
its corresponding velocity distribution (e.g. Tremaine~\cite{Tre01}). The pattern speed
then remains of the order of $\frac{M_d}{\sqrt{M_{\rm BH}}}$, for a sufficiently
cold disk, and is much smaller than the orbital frequency.  These slow waves
exist whenever the thin-disk Jeans length $\lambda_J = \frac{c^2}{G\Sigma_d}$
is lower than $4 r$, while the Toomre parameter $Q$ is less relevant 
(Lee \& Goodman~\cite{Lee99}). 

Although the waves observed in the N-body simulations were
non-linear, unwound and far from the WKB approximation, it is of
interest to compare the values of observed pattern speed with these 
predictions. Since our disk has significant velocity dispersion, the expected
modes are of the ``pressure'' nature, including the effects of dispersion
and gravity, corresponding to the $p$-modes of Tremaine (\cite{Tre01}).
We therefore expect positive pattern speeds.
If the effect of the background bulge can be ignored, our model simplifies
into a Kuzmin disk with a central point mass. It is more difficult to
determine the amount of equivalent softening, mimicking the effects of 
velocity dispersion. A reasonable approximation is that the Mach number
${\cal M}(r) = \Omega r /c \sim 3$ (see Fig.~\ref{omeg}). The reduction factor
 ${\cal F}(k c /\Omega)$ can be approximated as a decreasing exponential of
$k r \beta$, where $\beta \sim 1/{\cal M} \sim 0.3$. Fig.~5 from Tremaine (\cite{Tre01})
then predicts for the highest pattern speed (with the least nodes) a
value of $\Omega_p \sim 2.9$~km/s/pc, for a disk of 24\% the mass 
of the SBH (of $7\; 10^7$~M$_\odot$), almost coincident with what 
we observe in the simulations in Fig.~\ref{fig:pow_m19fpc} . This is a remarkable agreement,
given all the approximations.
Note that the Jeans length $\lambda_J$ maximises at $1.5 r$ 
over the disk.  

\subsection{Excitation of the waves}

 The waves can be maintained during a long time-scale, once 
excited, but triggering mechanisms remain to be found.
The displacement of the centre of mass (and corresponding
unstable oscillations) is not a good candidate, since the 
frequencies are much higher.

Tremaine (\cite{Tre95}) has
proposed that the $m=1$ wave amplifies through dynamical friction on the bulge, if
its pattern speed is sufficiently positive. This amplification results
from the fact that the friction decreases the energy less than the angular momentum.
The orbits with less and less angular momentum are more and more eccentric, and the
$m=1$ mode develops. To quantify numerically the dynamical friction of
the bulge would require a prohibitive number of particles in a live bulge.
We have attempted to compute the amplitude of the effect
through a semi-analytic method, as follows. During a simulation where
only the nuclear disk is represented by particles as above, the 
amplitude of a possible $m=1$ component, with its phase, are
computed through Fourier transform analysis at each 
radius, and each time step. The corresponding pattern speed
$\Omega_p$ is derived. The eccentricity of each orbit, 
averaged over 100 time-steps, is also derived. To the
particles participating in the $m=1$ (with the highest eccentricities),
is applied a dynamical friction force, as modelised
by Weinberg (\cite{Wei85}) for bars; this gives a torque proportional to
the pattern speed $\Omega_p$, since the bulge stars are assumed without
any significant rotation. The cumulative effects of the torques
over several thousands dynamical times were found negligible.
Note that in spiral galaxies like M~31, the bulge is slightly rotating,
which can counteract the effect on a slow pattern rotation.

Finally, an external perturbation is more likely to trigger the
$m=1$ perturbation: e.g. interstellar gas clouds are continuously
infalling onto the nucleus. Also, the slow modes pattern speed
discussed here are quite large with respect to the external
disk frequencies, and resonances are likely to occur.

\section{Conclusion}
\label{sec:conclusion}

In this paper, we have first tried to clarify a number of issues concerning the 
morphology and colours of the nucleus of M~31. The photometry in the nuclear
region of M~31 can thus be decomposed in 3 distinct components: 
\begin{itemize}
\item The bulge: a hot slowly rotating system with central colour values
of $(V-I)_B = 1.26$, $(U-V)_B = 1.98$ 
\item The nucleus: exhibiting two peaks in the $I$ band, the so-called P1 and P2, both
being offcentred with respect to the centre of the bulge isophotes.
Colours are suprisingly homogeneous throughout the nucleus
after the bulge contribution has been subtracted, confirming previous results,
as e.g. emphasized in KB99. However,
contrarily to the analysis of Lauer \etal\ (\cite{Lau98}), we found
that the nucleus is redder than the bulge in both $V-I$ and $U-V$ with
$(V-I)_N = 1.30$, $(U-V)_N = 2.36$.
\item A central blue excess, the so-called UV peak,
mostly apparent in the F300W/WFPC2 image,
where it is just resolved. This component was first detected 
and discussed by King \etal\ (\cite{Kin95}). It has an axis ratio of 
about $0.7$, a major-axis FWHM of $\sim 0\farcs21$ and a PA of $\sim 62\degr \pm 8$. 
\end{itemize}

We have then presented new \oasis\ and \stis\ data which complete 
our knowledge of the complex kinematics within the central 2'' of \MP. 
Taking the centre of the UV peak in the F300W WFPC2 image
as our reference zero point, we have derived the respective offsets 
required to reconcile
the FOC data of Sta+99, the SIS data of KB99,
and the newly presented STIS and \oasis\ kinematics.
The main results are summarized here:
\begin{itemize}
\item The kinematic axis (PA$=56.4\degr$) is nearly coincident with the major-axis
of the nucleus but not with the P1-P2 axis (PA$= 42\degr$).
\item The strong asymmetry of the velocity profile along the P1--P2 axis
observed by Sta+99 is confirmed. 
On the P1 side, the maximum
velocity in the \stis\ kinematics ($201\pm 5$~\kms) is reached at $0\farcs63$
(the turnover being at a radius of $0\farcs42$ with $V = 197\pm5$~\kms), 
while on the other side the minimum velocity($-292\pm21$~\kms) is at $-0\farcs235$.
\item The velocity dispersion peak is offset from the UV peak on the anti-P1 side.
It reaches $321\pm33$~\kms\  at $-0\farcs235$ in \stis\ measurements.
\item The velocity dispersion peak is consistent with being the result
of pure velocity broadening, and is not consistent with a hot system
located at the location of the UV peak. High velocities seem to be associated
with P2 which is observed to be offset from the UV peak on the anti-P1 side.
\end{itemize}

We finally performed new N-body simulations of $m=1$ modes
in a thin disk including a SBH in its centre. We have found that such
$m=1$ modes can be maintained from more than a thousand
dynamical times, when the disk mass represents between 20 and 40\%
of the central mass, allowing self-gravity to counteract
the differential precession of the disk. The pattern speed
of this $m=1$ mode is small, of the order of 3~km/s/pc in a
case designed to resemble M~31's nucleus. For the present
paper, we have only presented a model where the $m=1$
mode was excited by launching particles along elliptical orbits.
This model reproduces reasonably well the observed asymmetry in
the surface brightness distribution and mean stellar velocity, although
we had to remove some particles near the SBH to avoid numerical artefacts.
A detailed comparison with the observed stellar velocity dispersion 
requires a realistic treatment of the particles near the SBH.
The existence of an $m=1$ mode in the nucleus of M~31 is consistent
with the presence of a SBH with a mass in the range 
$\sim 3.5-8.5\, 10^7$ M$_\odot$.

The model proposed here relies on a relatively thin and cold nuclear
stellar disk. How can this disk be formed and maintained?
A likely possibility is that the central region is subject to accretion
and infall of gas from the nearby disk of M~31, which is gas rich. Already,
within the few 10-100~pc dust lanes, and CO molecular clouds are observed
(Melchior \etal\ \cite{Mel00}). The infall of gas clouds could both trigger the
$m=1$ instability, and also progressively reform the cold stellar disk.

\begin{acknowledgements}
We wish to warmly thank Catherine Dougados who accepted to carry out 
part of the \oasis\ observations reported in the present paper
using some of her allocated time at \cfht.
We thank the referee, Thomas Statler, for a detailed reading of the
paper and his helpful remarks.
We also thank Tim de Zeeuw for a critical reading of the manuscript. 
EE wishes to thank Richard Hook from ST/ECF at ESO for his help 
and comments regarding the {\tt drizzle2} routine.
Numerical computations have been carried out on the NEC-SX5 at IDRIS (Palaiseau, France).
\end{acknowledgements}

\appendix
\section{Ionised gas in the centre of \MP}

We have reexamined data obtained with the {\tt TIGER} spectrograph
(CFHT) in the spectral domain around the \NIIwa/\Haw\  emission lines, 
already discussed in Bacon \etal\ (\cite{Bac94}), 
to more systematically search for the possible presence
of ionised gas. The detection of a potential \Ha\ emission line
is difficult due to the underlying deep stellar \Ha\ absorption feature.
We have thus applied a new algorithm (Emsellem \etal, in preparation)
allowing an accurate subtraction of the stellar continuum, which exploits
a complete library of stellar and galaxy spectra with 0.5~\AA\ sampling.
The typical detection limit (3$\sigma$) is\footnote{Note 
that the detection limit quoted in Bacon~\etal\ (\cite{Bac94})
was wrong due to an error in the flux units.} then $3.6\,10^{-19}$~W.m$^{-2}$.

We are now successfully detecting an spectrally unresolved source of \Haw\ line emission,
about $2\farcs43$ ($0\farcs39$ North, $2\farcs4$ East)
from the centre of \M (Figs.~\ref{fig:emiline} and \ref{fig:mapline}). This emission
line was previously undetected mainly because it is well sunk into the 
corresponding absorption feature, The emission intensity peaks
at more than 5 times above the noise with a redshift of $-45\pm10$~\kms\  (heliocentric, 
see Fig.~\ref{fig:emiline}): this is to be compared
with the systemic velocity of $-308$~\kms, as derived from the OASIS spectra, which gives
a relative velocity of 263~\kms.
This emission line is observed on several adjacent spectra, 
its distribution being consistent with a spatially unresolved
system at the resolution of the {\tt TIGER} data (FWHM of $\sim 1\farcs1$). This is 
not an instrumental artefact, as neighbouring spectra come from non adjacent
regions of the CCD and exhibit the same spectral line. This is further supported
by the detection of a point-like source in 
the continuum subtracted F658N/\wfpc\ image (see Fig.~\ref{fig:mapline}),
as well as by recently published two-dimensional spectroscopy 
(del Burgo, Mediavilla and Arribas \cite{Del00}; source ``A'').
Note that the central sources A, B, C and D detected by del Burgo \etal\ (\cite{Del00})
all appear as point-like in the F658N/\wfpc\ image, their fluxes being consistent
with single planetary nebulae, in contradiction with the claims made by these authors.
\begin{figure}
\centering
\resizebox{5cm}{!}{\includegraphics[clip]{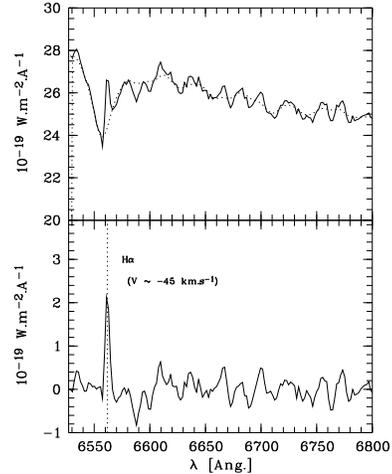}}
\caption{Top: original spectrum of \MP, summed over a region 
where \Ha\ emission is found (5 spectra). The dotted line shows the fit
performed using a library of high resolution spectra broadened and shifted
according to the local kinematics.
Bottom: resulting spectrum after subtraction of the template fit. The 
unresolved \Haw\ line clearly stands out, more than 5 sigma above the noise.}
\label{fig:emiline}
\end{figure}
\begin{figure}
\centering
\resizebox{6cm}{!}{\includegraphics[clip]{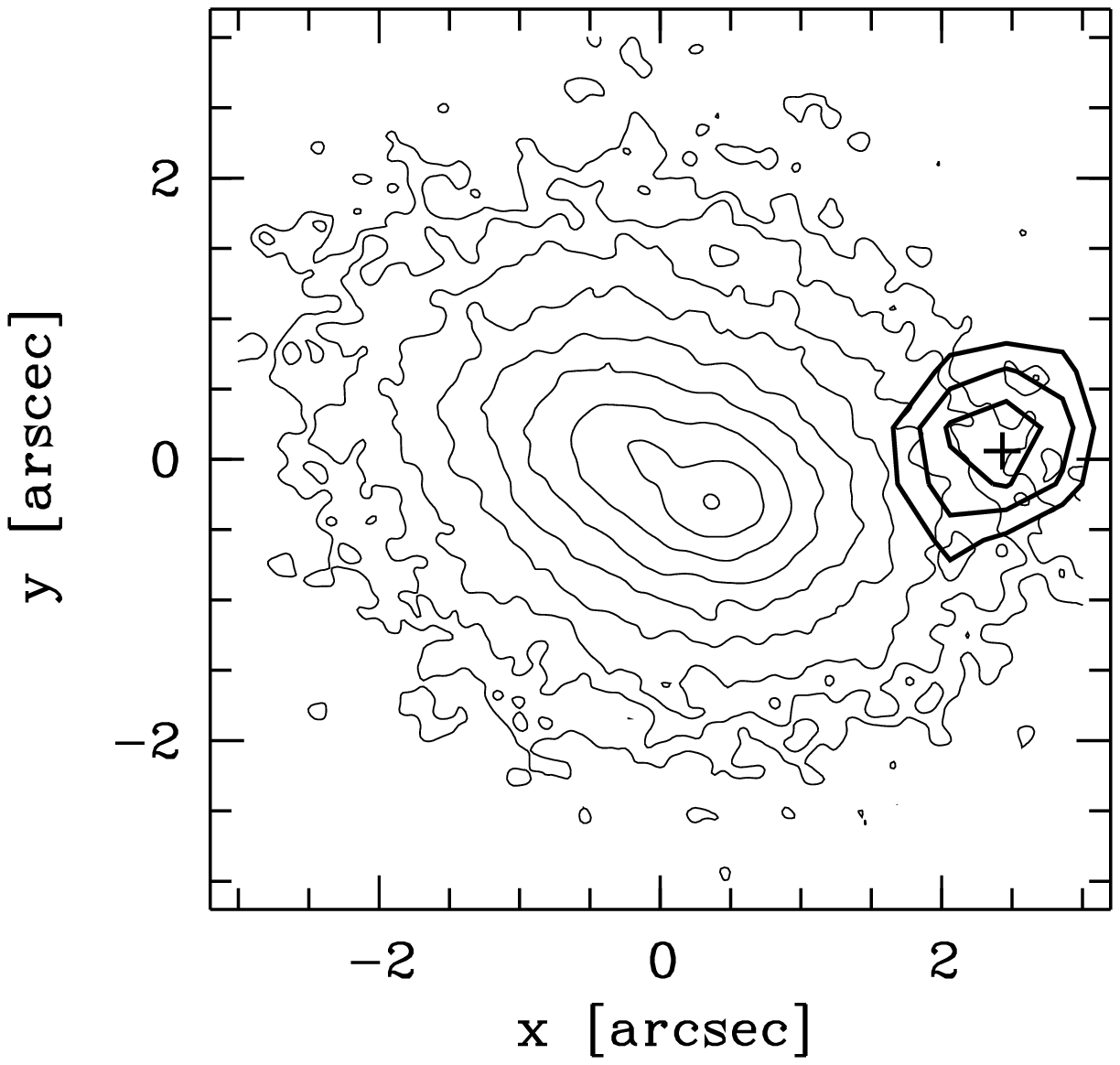}}
\caption{$I$ band \hst/\wfpc\ isophotes of the nucleus of \M (thin contours),
with the \Haw\ emission line flux contours overimposed ({\tt TIGER}, thick contours):
only isocontours more than 3 sigma above the noise have been included.
The cross marks the position of a point-like source detected
in the continuum subtracted \wfpc/F658N image.
The spatial sampling of the {\tt TIGER} data
was $0\farcs39$ per lens, and its resolution $\sim 1\farcs1$ FWHM.}
\label{fig:mapline}
\end{figure}

This feature can be simply explained by a single planetary nebula.
Indeed the integrated flux of this point-like source is $12.5\pm3\,10^{-19}$~W.m$^{-2}$
in the \wfpc/F658N continuum subtracted image (we derive a lower limit of $10\,10^{-19}$~W.m$^{-2}$ 
from the {\tt TIGER} datacube), compatible with a planetary in M~31.
The velocity of the detected \Ha\ emission line is consistent with the expected stellar velocities of
$\sim 250$~\kms\ at the edge of the nucleus on the P1 side, but is not consistent 
with the low velocities ($< 20$~\kms) of bulge stars. This strongly suggests that the 
object associated with this emission line belongs to the nucleus itself.
If we now assume that this emitting source is in the plane
of the nucleus, it is then at $\sim24\degr$ from the line of nodes, and
at $2\farcs81$ (or 10.5~pc) from the UV peak (for an assumed inclination of $i= 55\degr$).

\end{document}